\documentclass[twocolumn]{aastex631}
\usepackage{graphicx}	
\usepackage{amsmath}	
\usepackage{booktabs}
\usepackage{xtab}

\shorttitle{4U~1630--47}
\shortauthors{Fan et al.}

\begin{document}

\title{\textit{NICER} Spectral and Timing Analysis of 4U~1630--47 and its Heartbeat State}

\author{Ningyue~Fan}
\affiliation{Center for Astronomy and Astrophysics, Center for Field Theory and Particle Physics, and Department of Physics,\\
Fudan University, Shanghai 200438, China}
\affiliation{Center for Astrophysics, Harvard \& Smithsonian, Cambridge, MA 02138, USA}

\author{James~F.~Steiner}
\affiliation{Center for Astrophysics, Harvard \& Smithsonian, Cambridge, MA 02138, USA}

\author{Cosimo~Bambi}
\affiliation{Center for Astronomy and Astrophysics, Center for Field Theory and Particle Physics, and Department of Physics,\\
Fudan University, Shanghai 200438, China}
\affiliation{School of Natural Sciences and Humanities, New Uzbekistan University, Tashkent 100007, Uzbekistan}

\author{Erin Kara}
\affiliation{MIT Kavli Institute for Astrophysics and Space Research, MIT, 70 Vassar Street, Cambridge, MA 02139, USA}

\author{Yuexin~Zhang}
\affiliation{Center for Astrophysics, Harvard \& Smithsonian, Cambridge, MA 02138, USA}
\affiliation{Kapteyn Astronomical Institute, University of Groningen, P.O. BOX 800, 9700 AV Groningen, The Netherlands}

\author{Ole~K\"{o}nig}
\affiliation{Center for Astrophysics, Harvard \& Smithsonian, Cambridge, MA 02138, USA}

\correspondingauthor{Cosimo Bambi}
\email{bambi@fudan.edu.cn}

\begin{abstract}
We present a spectral and timing analysis of \textit{NICER} observations of the black hole X-ray binary 4U~1630--47 from 2018 to 2024. We find relativistic reflection features in the hard and soft intermediate states, and disk wind absorption features in the soft intermediate state and soft state. We fit the reflection features with \texttt{relxillCP} and find a stable and untruncated disk in the intermediate states; we fit the wind features with \texttt{XSTAR} and find a stable, highly ionized wind with high column density across different outbursts. Specifically, the heartbeat state is seen in two observations in 2021 and 2023 respectively. Through the phase-resolved spectral fitting, we find the flux of the source to be correlated with the disk parameters while no strong correlation with the coronal parameters is observed, consistent with the scenario given by the inner disk radiation pressure instability. A hard lag on the time scale of $1~\rm{s}$ and high coherence is observed near the characteristic frequency of the heartbeat, which can be explained by the viscous propagation of mass accretion fluctuations in the disk. The positive relationship between the heartbeat fractional rms and energy can possibly be explained by a disk-originated oscillation which is then magnified by the corona scattering.
\end{abstract}




\section{Introduction}

Black hole X-ray binaries (BHXRBs) are binary systems in which a central stellar-mass black hole accretes mass from a companion star. During an outburst, BHXRBs are usually divided into different states: the hard state, hard intermediate state (HIMS), soft intermediate state (SIMS) and soft state based on their spectral and timing properties \citep{Homan_Belloni_2005_statetransitions}. The hardness - intensity diagram \citep[HID,][]{Belloni_Motta_2011}, power color diagram \citep{Heil2015_powercolor} and hardness - rms\footnote{Referring to the fractional ``root mean squared'' (rms) noise. The definition of rms can be found in \cite{vanderklis1995}.} diagram \citep[HRD,][]{Plant2014_rms} are widely used in tracing the distinctive states of BHXRBs. Physically, different states are closely related to changes in the accretion process, as well as the outflows of the system in the form of winds and jets. The dynamic nature of the accretion process makes it possible to study the evolution of accretion geometry and the driving physics behind it \citep{Neilson2012_radiationpressure}.

How the disk-corona system changes structurally in different states is not clear yet. Spectral analysis is one of the useful ways to explore this question. In the hard state, the spectrum is dominated by non-thermal emission caused by
the Comptonization of thermal photons off hot electrons in the corona, and the reflection of those Comptonized photons from the disk.
In the soft state, thermal emission from the accretion disk is dominant. Through the modeling and fitting of different spectral components, we can analyze the properties and evolution of the system from the spectral parameters \citep[e.g.,][]{Wang2018_gx339,2021DeMarco_1820,Liu2023_gx339,Fan_maxij1820}.

Besides spectral features, studying variability on different time scales enables us to investigate the innermost region of the accretion flow \citep{Ingram2019_QPOreview}. Low-frequency (LF) quasi-periodic oscillations (QPOs) with a centroid frequency in the range of 0.1-30~Hz are common in BHXRBs. The power density spectrum (PDS, derived from the Fourier transform of the light curve) is a useful way to study this behavior. Based on different features of the PDS, LFQPOs are divided into types-A, B and C \citep{Casella2005_ABC}, which appear in different states of the outburst \citep{Sriram,Ma2021_highenergyQPO}.

One characteristic QPO-like feature of particular note appears in the ``heartbeat” state, for which  an electrocardiogram-like light curve modulates quasi-periodically at a much lower frequency around $\sim 10$~mHz \citep{Neilsen2011_physicsofheartbeat}.  The low-frequency high-amplitude oscillation is distinctive from other QPOs and is also sometimes referred to as a quasi regular modulation \citep[QRM,][]{Trudolyubov2001_1998qrm,Yang2022_qrm2021}.  The heartbeat state has only been seen in a handful of BHXRBs, including GRS 1915+015 \citep{Neilsen2011_physicsofheartbeat,Neilson2012_radiationpressure}, IGR~J17091--3624 \citep{Wang2024_IGRJ17091,Katoch2021_IGRJ17091_GRS1915}, GRO J1655--40 \citep{Remillard1999_GROJ1655}, and 4U~1630--47 \citep{Trudolyubov2001_1998qrm,Yang2022_qrm2021} (on which we focus).

The underlying  mechanism behind the heartbeat is not fully understood. Commonly, the heartbeat mechanism is thought to be associated with a radiation pressure instability of the inner disk \citep{Lightman1974_instability,Done2007review,Neilson2012_radiationpressure}, a hypothesis which is broadly supported by phase-resolved and time-resolved spectral fitting \citep{Neilsen2011_physicsofheartbeat,Rawat2022_GRS1915_timeresolved,Belloni1997_time}. Moreover, the time scale of the modulation is found to be commensurate with the viscous timescale of the inner disk region \citep{Belloni1997_time}. \cite{Wang2024_IGRJ17091} further discussed the possibility of the heartbeat arising from the vertical disk instability in systems with a long orbital period \citep{Zycki1999} and disk tearing due to spin-orbit misalignment \citep{Raj_Nixon_2021_disktearing} in the case of IGR~J17091--3624.

4U~1630--47 is a recurrent transient which typically outbursts every few hundred days \citep{Choudhury2015,Jones1976}.
Previous spectral and timing studies on 4U~1630--47 have shown interesting features of this source: It has
a high hydrogen column density along the line of sight, estimated to be of the order $N_{\rm H} \approx 10^{23}\,\mathrm{cm^{-2}}$ with high resolution \textit{Chandra} data \citep{Gatuzz2019_chandrahighsolution}. Its dimensionless spin parameter has been estimated to be higher than 0.9 through both continuum fitting and reflection modeling \citep{Kushwaha2023_HSS,kING2014_incl,Pahari2018_spin_wind}, but see also \citet{Qichun2022}. The inclination angle is measured to be quite high, around $\sim64^{\circ}
$ through reflection modeling \citep{kING2014_incl}. Prominent absorption features from disk winds are commonly seen for 4U~1630--47, e.g., in \cite{Pahari2018_spin_wind,Trueba2019_4U1630wind,Kushwaha2023_HSS, Ratheesh2024_HSS}. The source is mainly observed in its soft state. Through spectral and timing analysis of \textit{INTEGRAL} and \textit{RXTE} data, \cite{2015Capitanio_missinghard} find that the hard state is missing in the 2006 and 2008 outbursts. The heartbeat state of this source was first observed in 1998 with \textit{RXTE} \citep{Trudolyubov2001_1998qrm}. \cite{Yang2022_qrm2021} observe a clear transition from a hard-state with a low-frequency type-C QPO to the (lower-frequency) heartbeat state in 2021 with \textit{Insight}-HXMT. Notably, for the two heartbeat states of 4U~1630--47 observed previously, the level of luminosity and hardness is similar \citep{Yang2022_qrm2021}. 

In this paper, we focus on the spectral and timing analysis of \textit{NICER} observations of 4U~1630--47 over the past six years. Since 2018, \textit{NICER} has observed this source through several outbursts, and at a high monitoring cadence (see Fig.~\ref{lightcurve}), which enables detailed assessment of the common patterns in its behavior.

The paper is organized as follows. Section~\ref{observation} provides a detailed account of the observations and data reduction process, while Section~\ref{fitting} presents the applied spectral models and the corresponding results. The phase-resolved fitting of the heartbeat state is presented in Section~\ref{phase fitting} and Section~\ref{timing} shows the energy-resolved timing analysis of the heartbeat state. The discussion is presented in Section~\ref{discussion}, and the conclusion is in Section~\ref{conclusion}.

\section{Observation and Data Reduction}\label{observation}

\begin{figure*}
    \centering
    \includegraphics[width=0.98\linewidth]{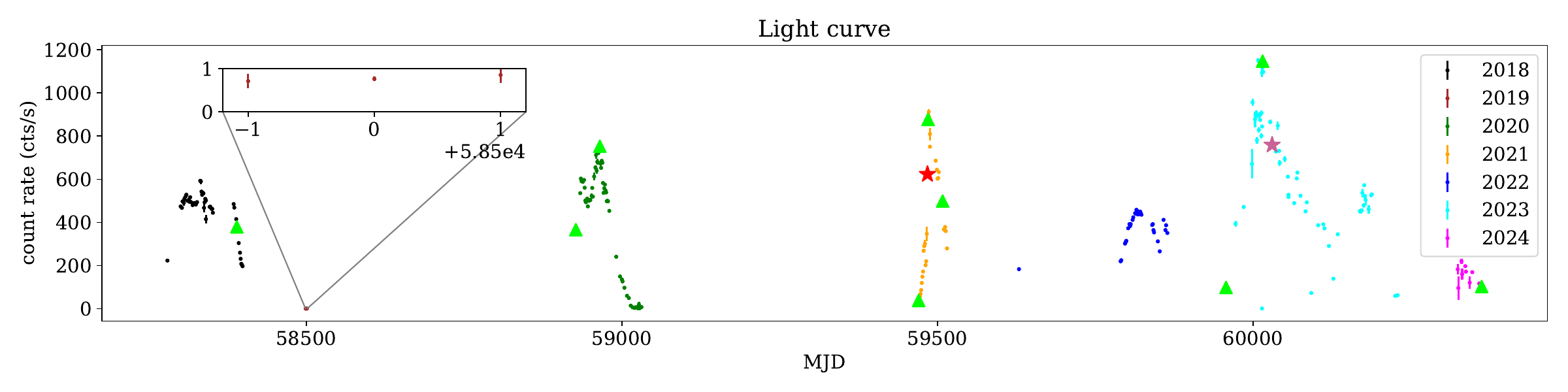}
    \caption{Light curve of the \textit{NICER} observations of 4U~1630--47 from 2018 to 2024. Observations are colored by calendar year. The two stars mark the observations where the heartbeat is seen. Triangles mark the observations used to constrain the elemental abundances with \texttt{tbfeo} in Section~\ref{fitting}.}
    \label{lightcurve}
\end{figure*}

\begin{figure}
    \centering
    \includegraphics[width=0.98\linewidth]{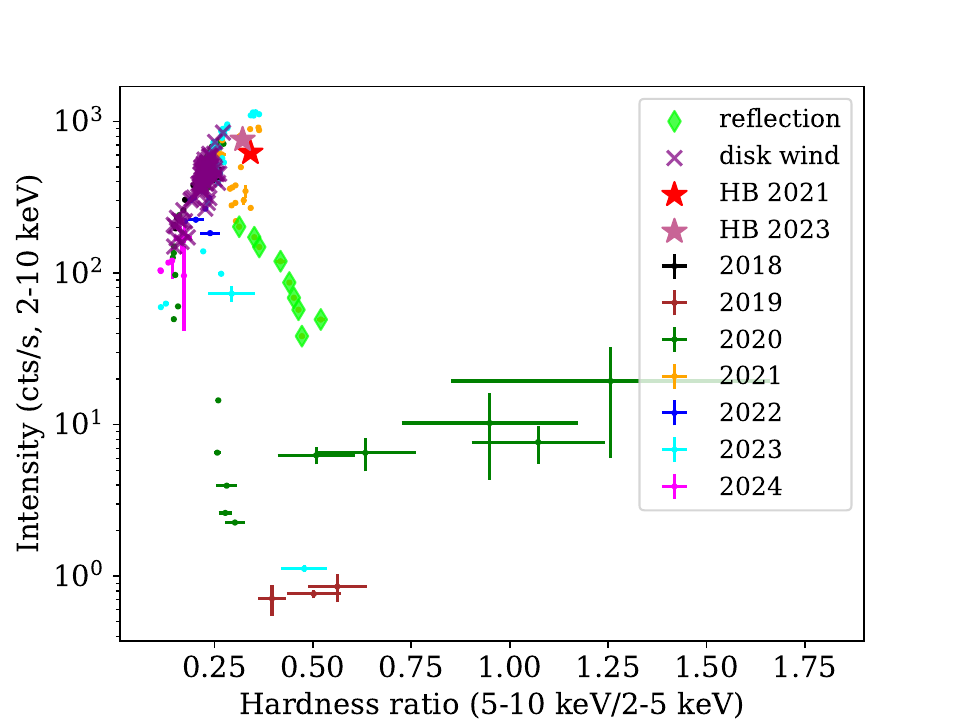}
    \caption{The HID of the \textit{NICER} observations of 4U~1630--47 from 2018 to 2024. The hardness is defined as the ratio between the count rates in the 5--10~keV band and the 2--5~keV band. Observations are color-coded as in Fig.~\ref{lightcurve}. Stars mark the observations where the heartbeat is seen. Purple crosses mark the spectra with clear disk wind absorption features, which are observed in numerous outbursts, including 2018, 2020, 2022, 2023 and 2024. Green diamonds mark observations with pronounced reflection features, which all come from in the beginning of the 2021 outburst.}
    \label{HID}
\end{figure}

\begin{figure}
    \centering
    \includegraphics[width=0.98\linewidth]{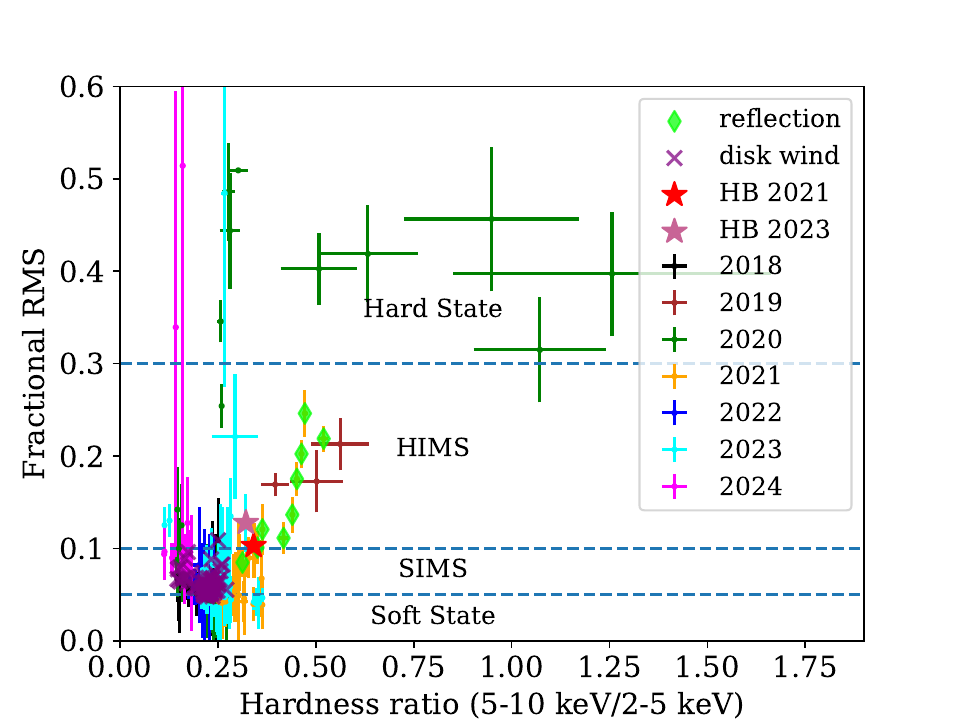}
    \caption{The HRD of the \textit{NICER} observations of 4U~1630--47 from 2018 to 2024. The fractional rms is computed as the unbiased standard deviation of the broadband light curve divided by the mean. Color and symbol definitions are the same as in Fig.~\ref{HID}. The blue dashed lines show the boundary of different states, based on the classification in \cite{Plant2014_rms}.}
    \label{frms-hardness}
\end{figure}

We analyze 251 \textit{NICER} observations of the source from 2018 to 2024. During these years, the \textit{NICER} calibration is not known to have changed appreciably. But in May 2023, there was an increase in optical loading on the instrument, and as a result, during the ISS daytime, increased noise can affect the study of low-energy photons below 1~keV\footnote{https://heasarc.gsfc.nasa.gov/docs/nicer/analysis\_threads/light-leak-overview}. However, this is unlikely to influence our study on 4U~1630--47, because only higher energy photons are considered in spectral and timing analysis due to the high interstellar absorption.

We perform the standard calibration and screening on the archival \textit{NICER} data with \texttt{nicerl2}\footnote{https://heasarc.gsfc.nasa.gov/docs/nicer/analysis\_threads/nicerl2/} in \texttt{HEASOFT v6.31.1}. We use the \texttt{3C50} background model  \citep{Remillard2022_3C50}. We then use the \texttt{nicerl3-spect}\footnote{https://heasarc.gsfc.nasa.gov/docs/nicer/analysis\_threads/nicerl3-spect/} task to generate the source and background spectra, and the corresponding response files for each observation. The public software \texttt{stingray}\footnote{https://docs.stingray.science/en/stable/index.html} v2.1 is used to generate the light curve in the 0.2--12~keV energy band (Fig.~\ref{lightcurve}), the hardness-intensity diagram (HID, Fig.~\ref{HID}), and the PDS of the light curves from the original events files. Throughout the paper, uncertainties are reported at the 90\% confidence level. The count rates of different detectors are normalized to 52-FPM equivalent intensity, and we adopt this normalization throughout this work. When computing the PDS, the time resolution of the light curves is 0.005~s, corresponding
to a Nyquist frequency of 100~Hz. To create an averaged PDS from different segments of a light curve, the size of each segment is chosen to be 256~s, corresponding to a minimum frequency of about 0.003~Hz, which is enough for the study of heartbeat features at several mHz.

From the light curve, we can see that in the observations in most years, as a new outburst onsets, the count rate first rises over a period of a few weeks and later decreases over a timeframe which can be similar or much longer. Since the HID of this source does not follow the ``Q''-shaped track typical of most BHXRB transients, and additionally complicated by the very large column-density in the line of sight, it is challenging to distinguish different states from the HID in isolation. Therefore, we additionally explore the timing-variability features as well as the spectral features, and we show the hardness-rms diagram (HRD) in Fig.~\ref{frms-hardness}. Based on classifications of states discussed in \cite{Plant2014_rms} for GX 339--4 with \textit{RXTE}, \cite{Alabarta2020_1727HRD} for MAXI J1727--203 observed by \textit{NICER}, \cite{Stiele2020_1820HRD} for MAXI J1820 observed by \textit{NICER} and \textit{Swift}, and more general discussions in \cite{Belloni2010,Belloni2016}, we show the distinction of different states based on the level of rms on the HRD. We find most of the observations in this 6 year-interval are in the SIMS or soft state. However, in 2021, we detect a clear transition from the HIMS to SIMS.

\begin{figure}
    \centering
    \includegraphics[width=0.98\linewidth]{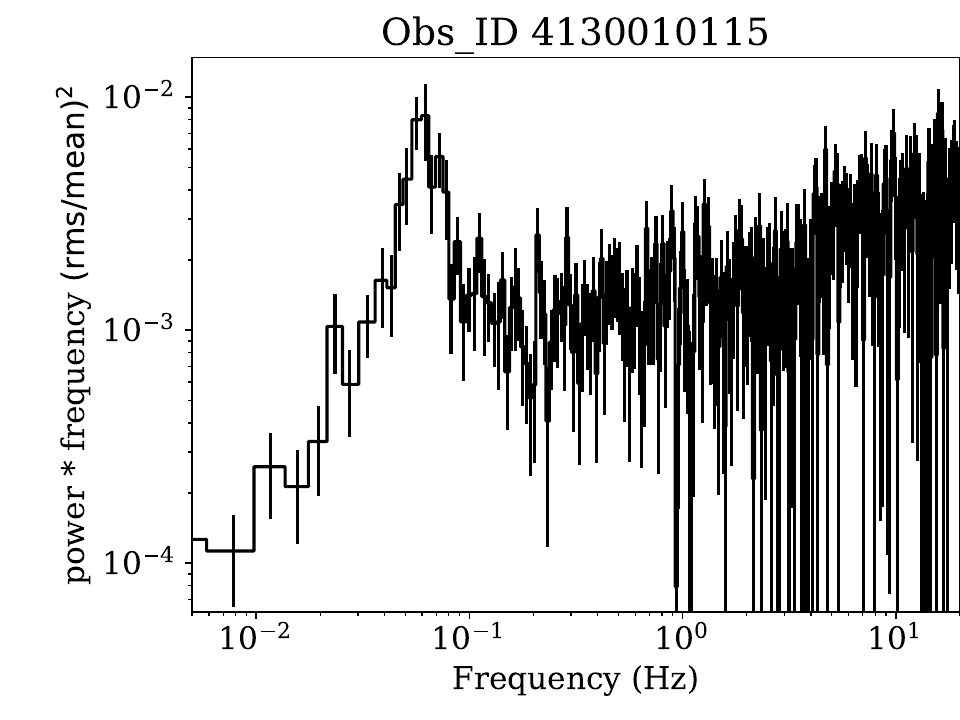}
    \includegraphics[width=0.98\linewidth]{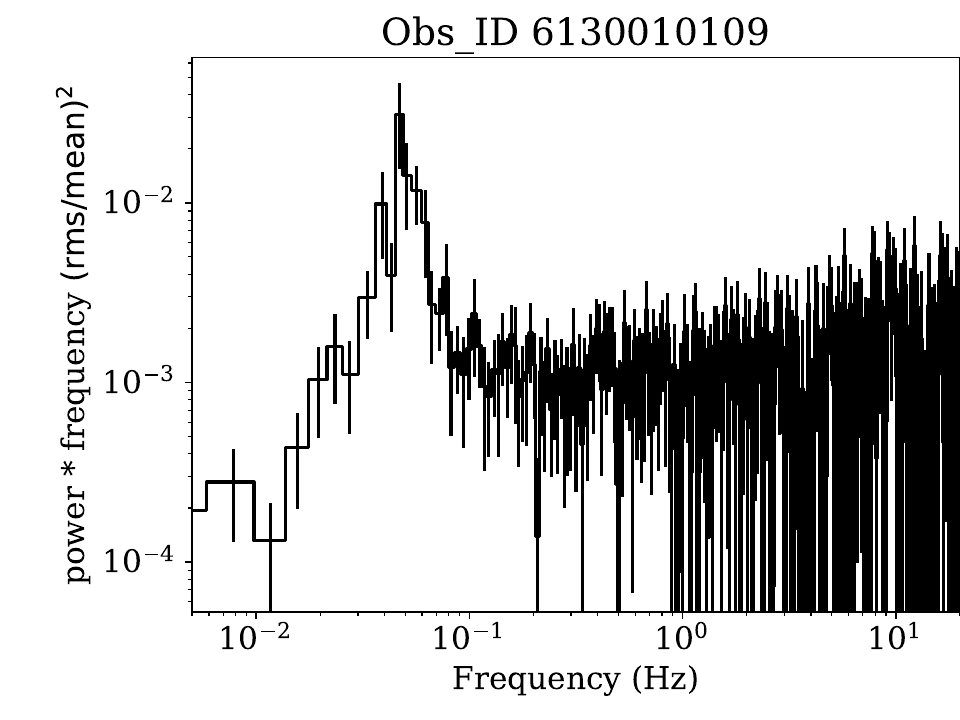}
    \caption{The PDS of the two observations when the heartbeat is observed. Observation ID 4130010115 corresponds to the observation on September 27, 2021 (MJD 59485), while Observation ID 6130010109 corresponds to the observation on March 27, 2023 (MJD 60031). For the 2021 heartbeat, the PDS peaks at $5.6^{+0.4}_{-0.6}$~mHz. For the 2023 heartbeat, the PDS peaks at $4.7^{+0.4}_{-0.4}$~mHz.}
    \label{PDS}
\end{figure}

\begin{figure}
    \centering
    \includegraphics[width=0.98\linewidth]{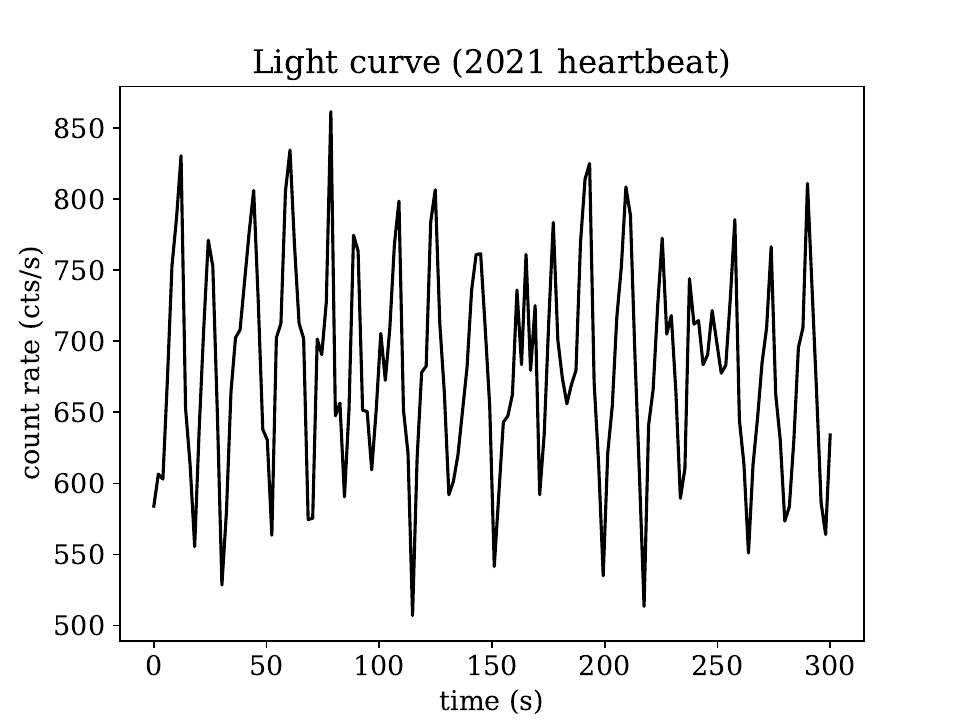}
    \includegraphics[width=0.98\linewidth]{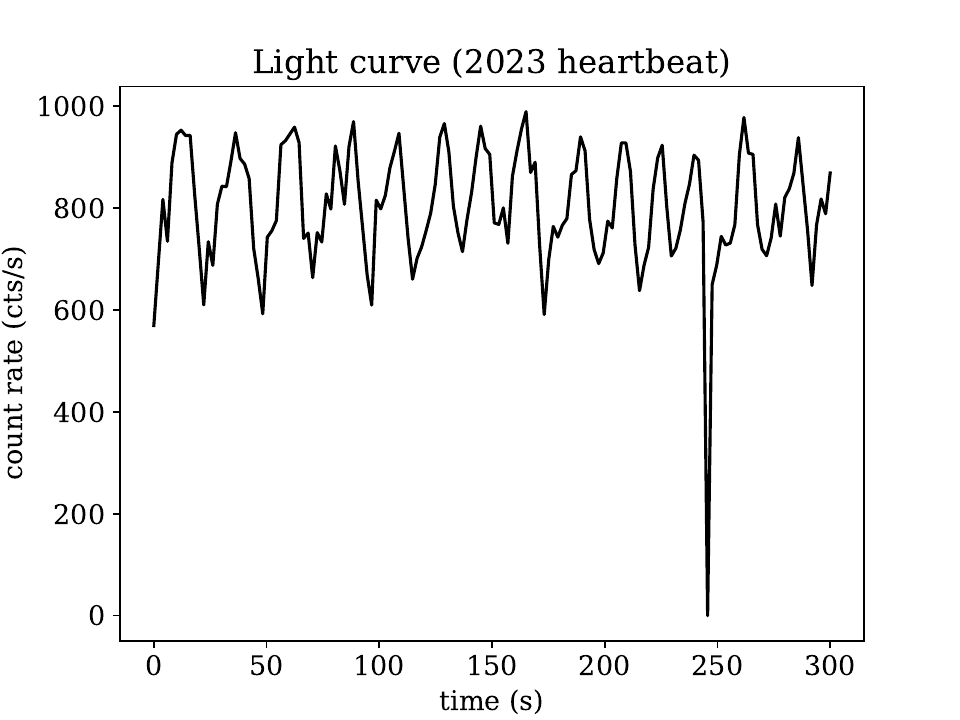}

    \caption{The 300~s light curve segments of the heartbeat states corresponding to the PDS in Fig.~\ref{PDS}. The upper panel is the 2021 heartbeat state, with an oscillation period of about 18~s (measured from phase-folding). The lower panel shows the 2023 heartbeat state, with an oscillation period of about 20~s (measured from phase folding).}
    \label{phase_lc}
\end{figure}

Among all the data obtained by {\em NICER}, only two observations (on September 27, 2021 (MJD 59485, ObsID 4130010115) and March 27, 2023 (MJD 60031, ObsID 6130010109) respectively) show strong heartbeat features. Their PDS (Fig.~\ref{PDS}) show a clear,  broad peak (compared to a LFQPO) at around 5~mHz, corresponding to a modulation period of around 20~s. This modulation pattern can also be clearly identified in the raw light curves of these two observation (Fig.~\ref{phase_lc}). As is shown in Fig.~\ref{HID} and Fig.~\ref{frms-hardness}, the heartbeat states of 4U~1630--47 are localized to a precise phase of the outburst, i.e., at a particular locus in luminosity, hardness and rms-timing variability and that locus is at the point of HIMS-to-SIMS transition.

\section{Spectral analysis and Results}\label{fitting}

Spectral fitting is conducted with \texttt{XSPEC} v12.13.0 \citep{Xspec_Arnaud}. We use the elemental abundances of \cite{Wilms_tbabs} and cross-sections of \cite{Verner_crosssection}. The $\chi^{2}$ statistics is used to find the best-fit values and uncertainties (throughout the paper given at a 90\% confidence level) of the parameters. In the fits, we analyze the spectra in the 2--10~keV energy range.  Owing to the high interstellar absorption, energies below this are dominated by the response-shelf. 

\subsection{Simple continuum model}

We first fit the spectra with a simple model: \texttt{tbabs $\times$ diskbb}. We employ the \texttt{diskbb} model \citep{Mitsuda1984_diskbb} to fit the multi-temperature blackbody component from the accretion disk and the \texttt{tbabs} model to account for interstellar absorption \citep{Wilms_tbabs} with the column density $N_{\rm H}$ as a free parameter in the fits. We find $N_{\rm H} \approx 10^{23}\,\mathrm{cm^{-2}}$, consistent with earlier studies \citep{Gatuzz2019_chandrahighsolution,Pahari2018_spin_wind}. While a significant fraction of the observations are successfully fitted with this simple model (i.e., with a reduced $\chi^{2}/\nu \sim 1$), most of the fits yield unacceptable fits with $\chi^2/\nu$ even reaching 3-4 in some cases. This indicates that most of the spectra of this source cannot be described well by this simple model, therefore we need to adapt more complex models and take other spectral components like the Comptonization component from the corona into consideration.

\begin{figure}
    \centering
    \includegraphics[width=0.98\linewidth]{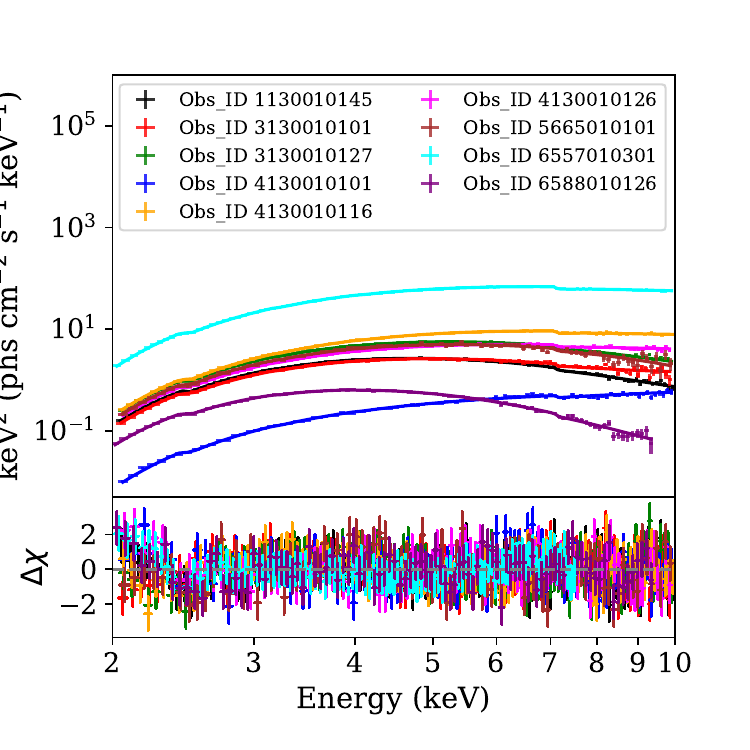}
    
    \caption{Fits and their residuals for the selected spectra shown with green triangles in Fig.~\ref{lightcurve} with fits using the model \texttt{tbfeo $\times$ thcomp $\otimes$ diskbb}. In these fits, the \texttt{tbfeo} parameters are linked across the spectra to obtain a robust constraint leveraging spectra with different spectral shapes.}
    \label{tbfeo}
\end{figure}

\begin{figure*}
    \centering
    \includegraphics[width=0.98\linewidth]{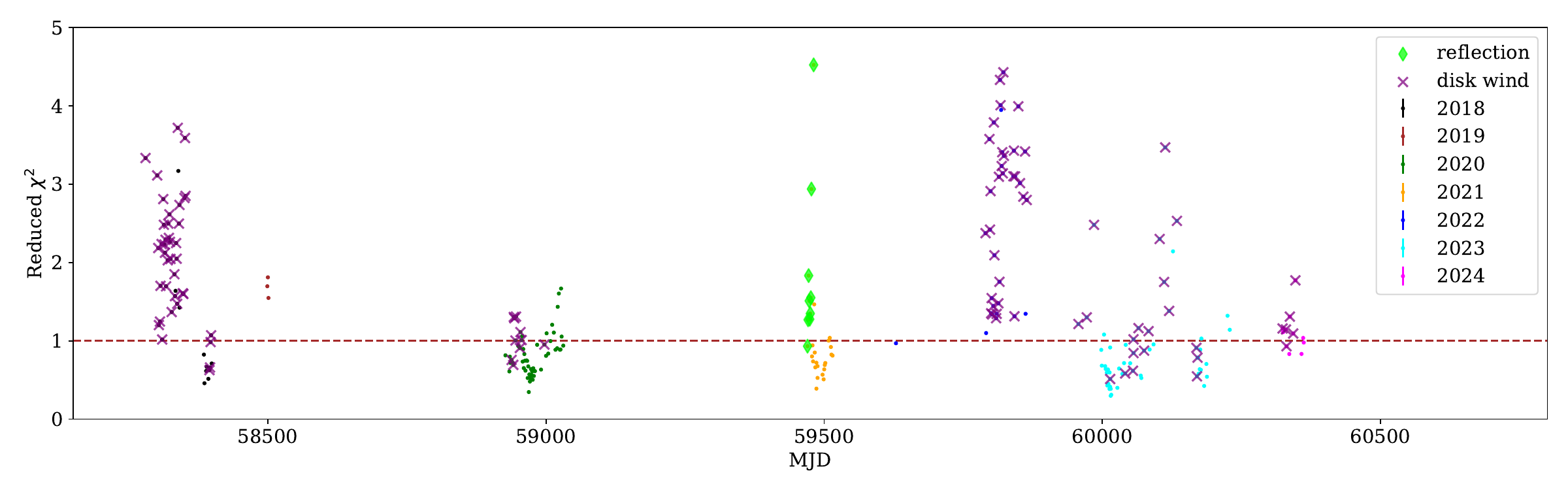}
    \caption{The reduced $\chi^{2}$ of the fits with the model \texttt{tbfeo $\times$ thcomp $\otimes$ diskbb}. The purple crosses mark the spectra with clear disk wind absorption features. The green diamonds mark the spectra with reflection features.}
    \label{chi2_thbb}
\end{figure*}

\begin{figure}
    \centering
    \includegraphics[width=0.98\linewidth]{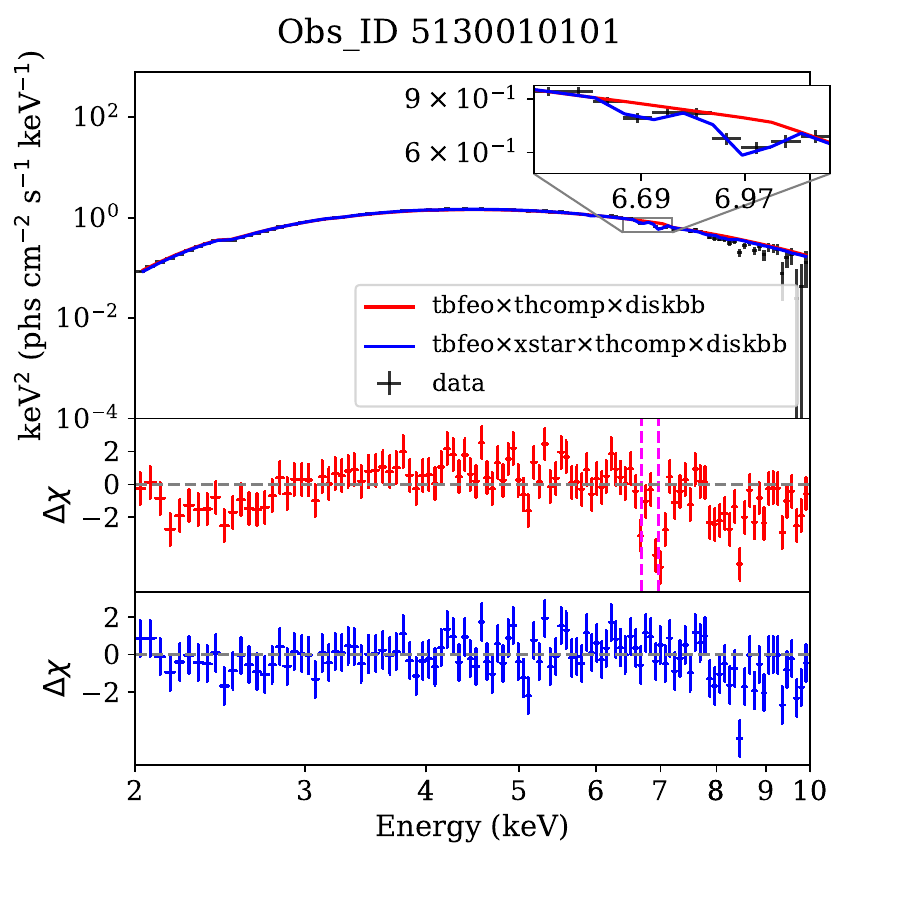}

    \caption{A typical and representative spectrum with disk wind features, taken from Observation ID 5130010101 obtained on July 30, 2022 (MJD 59786). The top panel shows the fits with (blue) and without (red) having included a wind component. The two magenta dashed vertical lines mark the 6.69~keV Fe XXV line and 6.97~keV Fe XXVI line. The two bottom panels show the residuals from both fits.}
    \label{wind}
\end{figure}

\begin{figure}
    \centering
    \includegraphics[width=0.98\linewidth]{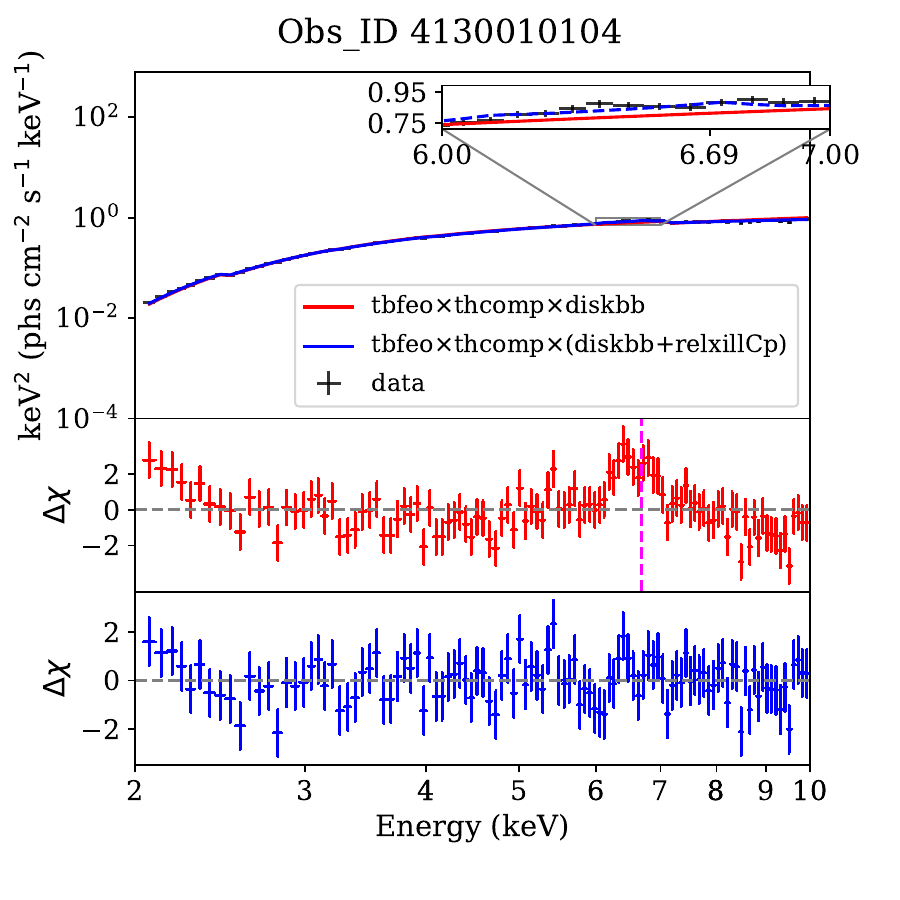}

    \caption{A typical and representative spectrum with relativistic reflection features, taken from Observation ID 4130010104 obtained on September 16, 2021 (MJD 59475). The top panel shows the fits with (blue) and without (red) having included a reflection component. The magenta dashed vertical line marks Fe XXV at 6.69~keV. The bottom panel shows the fitting after the reflection component \texttt{relxillCp} is added.}
    \label{reflection}
\end{figure}

We then fit the spectra with the model \texttt{tbfeo $\times$ thcomp $\otimes$ diskbb}, where \texttt{thcomp} \citep{Zdziarski2020_thcomp} is a convolutional model used to fit the Comptonization of thermal photons generated by the inverse Compton scattering of thermal photons off free electrons in the corona, and \texttt{tbfeo} is an interstellar absorption model which allows modification of the oxygen and iron abundances in addition to the hydrogen column density. 

To determine the hydrogen column density and elemental abundances in \texttt{tbfeo}, which should be stable over time, we select nine observations spanning different accretion rates and states (three with low count rate, three intermediate, and three with high count rate.  These are indicated by the green triangles in Fig.~\ref{lightcurve}). We fit these simultaneously with the column density and elemental abundances in \texttt{tbfeo} tied together while the disk and coronal parameters are left free for each observation. The joint fit (shown in Fig.~\ref{tbfeo}) gives a reduced $\chi^{2}$ of 743/937 = 0.8 (perhaps indicating that the systematic uncertainties adopted for the {\em NICER} processing are overestimated). From this, we obtain a column density of $N_{\mathrm{H}}=16.96^{+0.58}_{-0.41}\times10^{22}~\mathrm{cm^{-2}}$, and oxygen and iron abundances of $[O]=0.21^{+0.05}_{-0.06}$ and $[Fe]=0.69^{+0.07}_{-0.05}$, respectively. This fit is shown in Fig~\ref{tbfeo}, from which it can be seen that the model tracks our reference spectra without any obvious outlier observations performing poorly. In our subsequent analysis, we fix the parameters of \texttt{tbfeo} at the values obtained here. However, we note that only fitting data above 2~keV limits our ability to robustly constrain individual elemental abundances.  One can reasonably expect that O abundance fitted acts as a proxy for the aggregate of low-Z metals.  

The reduced $\chi^{2}$ of the fits with the model \texttt{tbfeo $\times$ thcomp $\otimes$ diskbb} is shown in Fig.~\ref{chi2_thbb}. While the reduced $\chi^{2}$ of some of the fits decreases to around one, indicating that the new model is much more successful, we identify numerous outliers for which the fit remains unacceptable. The residuals of these fits reveal in several cases, clear wind absorption features (observations marked by purple crosses in Fig.~\ref{chi2_thbb}), whereas others exhibit relativistic disk-reflection features (marked by green diamonds in Fig.~\ref{chi2_thbb}). Further details on the quantitative definition of wind features are given in Appendix~\ref{app:A}.  

The typical spectra with disk wind features and relativistic reflection features are shown in the top panels of Fig.~\ref{wind} and Fig.~\ref{reflection}, respectively. For disk wind features, there are two absorption lines between 6--7~keV (Fe XXV line at 6.69~keV and Fe XXVI line at 6.97~keV). For reflection features, a broadened iron K$\alpha$ emission line around 6.69~keV is observed. In the following two subsections, we will further discuss how we improve the fits of these spectra with disk wind absorption and relativistic disk reflection features.

\subsection{Disk wind absorption features}

\begin{figure}
    \centering
    \includegraphics[width=0.998\linewidth]{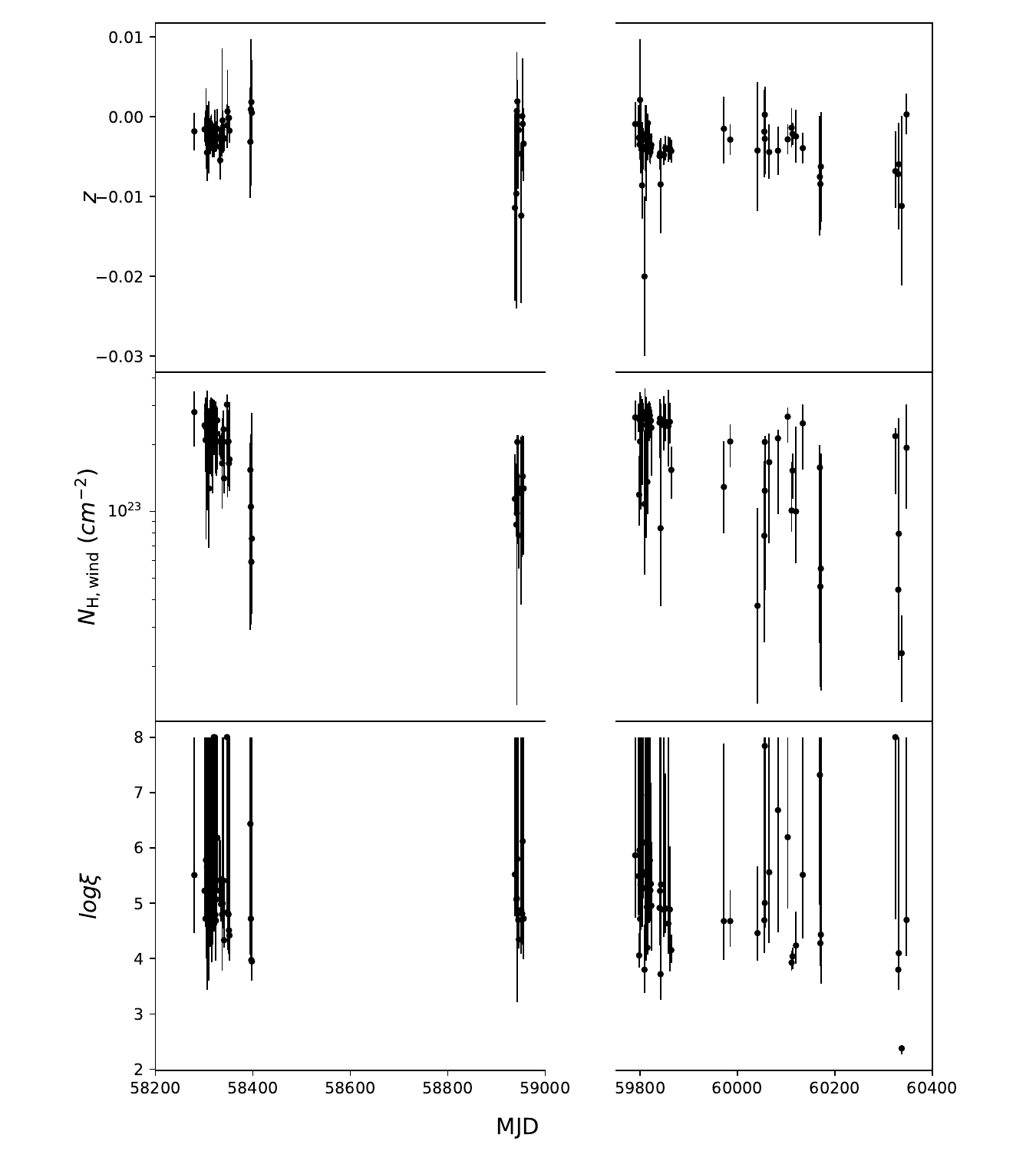}
    \caption{Fitting results of the parameters in the model \texttt{tbfeo $\times$ xstar $\times$ thcomp $\otimes$ diskbb}. The parameters are (from top to bottom): the red shift, the column density, and the ionization degree of the wind.}
    \label{wind fit}
\end{figure}

We calculate a grid model with \texttt{XSTAR} \citep{Kallman2001_xstar} and use the table model to fit the absorption by an ionized wind. The input continuum spectrum of the \texttt{XSTAR} calculation is the averaged best-fit spectrum with the continuum model. Becuase of the similar thermal properties of all the spectra showing wind absorption features (see Fig.~\ref{frms-hardness}), almost all spectra can be fit well with the \texttt{XSTAR} table model with the same seed spectrum. The model we adopted to fit these spectra is: \texttt{tbfeo $\times$ xstar $\times$ thcomp $\otimes$ diskbb}, and the three free fitting parameters include the absorber column density ($N_{\rm H, wind}$), the ionization degree ($\log\xi$), and the redshift ($z$). The fitting results are shown in Fig.~\ref{wind fit}. From year-to-year, no significant evolution is apparent in the properties of the wind. On average, the column density of the wind is $N_{\rm H, wind} \approx (1.9\pm0.9)\times10^{23}\,\mathrm{cm^{-2}}$, the ionization parameter $\log\xi \approx 5\pm2$, and the redshift is $z \approx (-0.003\pm0.002)$.

\subsection{Relativistic reflection features}

We use the relativistic reflection model \texttt{relxillCp} \citep{Garcia_relxill} to fit the the reflection features. From 2018 to 2024, there are only 9 \textit{NICER} observations showing a clear reflection component, which are observed during the HIMS to SIMS state transition at the beginning of the 2021 outburst. The model used to fit these spectra is \texttt{tbfeo $\times$ thcomp $\otimes$ (diskbb + relxillCp)}. To avoid the degeneracy between the inner disk radius and the spin \citep{Buisson2019}, we fix the spin at the maximum value 0.998 in \texttt{relxillCp} and study the change of $R_{\mathrm{in}}$. The inclination angle is fixed at $64^{\circ}$ \citep{kING2014_incl}. The reflection fraction is set to $-1$ so that only the reflection component is produced (noting that the Comptonized emission is already described by \texttt{thcomp $\otimes$ diskbb}). The coronal temperature $kT_\mathrm{e}$ and photon index $\Gamma$ in the Comptonization component and reflection component are linked together. Due to the lack of data higher than 10~keV, the corona temperature cannot be constrained well from the fits, so we fix it at a reasonable benchmark of 50~keV. We also checked that a different corona temperature does not significantly impact the fit results. For example, when we fix the corona temperature at 100~keV instead, only the absolute values of the photon index $\Gamma$ become slightly lower, while the trends and values of other parameters do not change much. The emissivity of the reflection component is modeled with a power law ($\epsilon \propto r^{-q}$).

\begin{figure}
    \centering
    \includegraphics[width=0.85\linewidth]{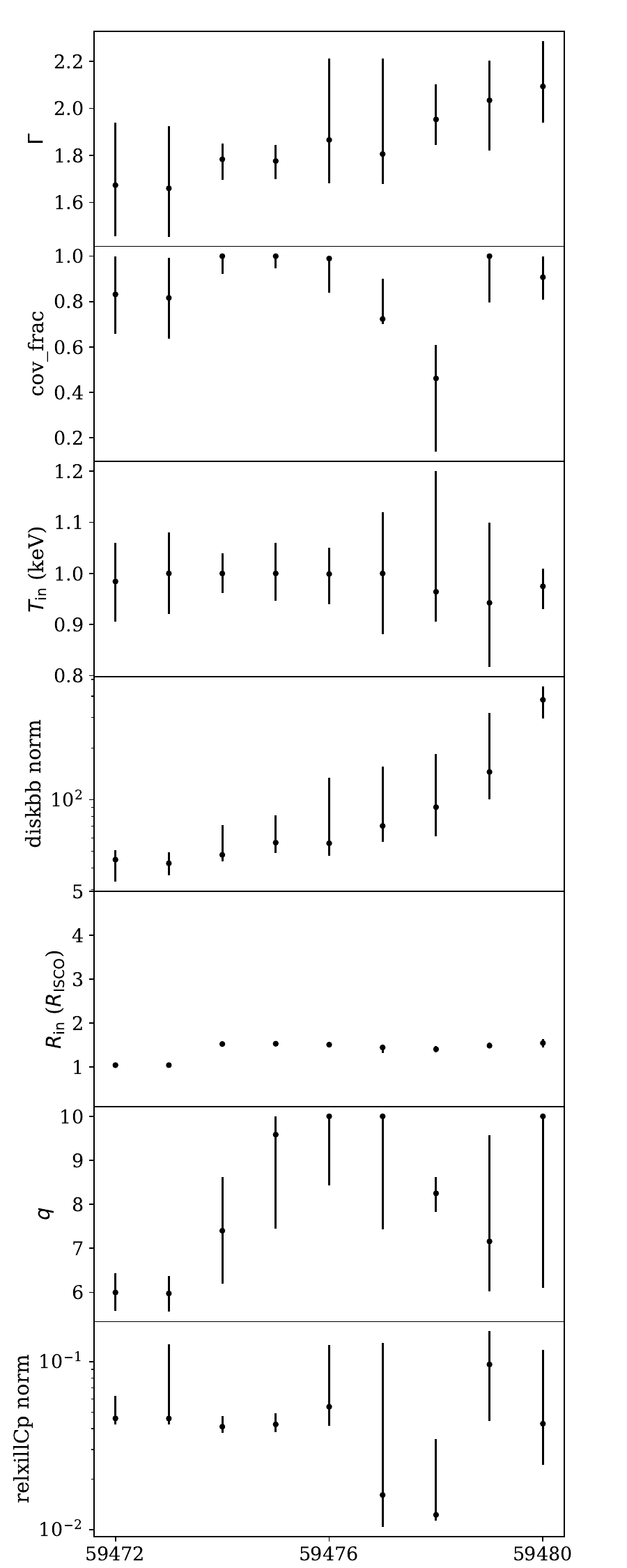}

    \caption{Fitting results of the parameters in the model \texttt{tbfeo $\times$ thcomp $\otimes$ (diskbb + relxillCp)}. The parameters are (from top to bottom): the photon index $\Gamma$, the covering fraction of the corona over the disk, the inner disk temperature, the normalization factor of the disk component, the inner disk radius in units of $R_{\mathrm{ISCO}}$ for a Kerr black hole with $a_{*}=0.998$ given by \texttt{relxillCp}, the reflection emissivity index $q$, and the normalization of the reflection component.}
    \label{reflection fit}
\end{figure}

The fitting results for these nine spectra are shown in Fig.~\ref{reflection fit}, from which we can see how the disk and corona evolve during the state transition from the HIMS to SIMS. The photon index $\Gamma$ changes from around 1.7 (which is the typical value of $\Gamma$ in the harder states) to higher than 2 (which is the typical value of $\Gamma$ in the softer states). For the thermal component, while the inner disk temperature does not change much, the normalization factor of \texttt{diskbb} increases. Concerning the ongoing debate on whether the disk is truncated or not in the hard state and intermediate states \citep{determ_spin_truncation}, our results here indicate that the inner disk radius $R_{\mathrm{in}}$ given by the reflection component is stable and near the innermost stable circular orbital (ISCO) in the HIMS to SIMS transition.

Meanwhile, in the first 7 observations in the HIMS, an increasing trend of the reflection emission index $q$ is observed while the covering fraction of the corona is decreasing, indicating a possible scenario that the reduced spatial extent of the corona makes the reflection emission more concentrated centrally.

\subsection{Fitting results of the thermal component}

\begin{figure}
    \centering
    \includegraphics[width=0.98\linewidth]{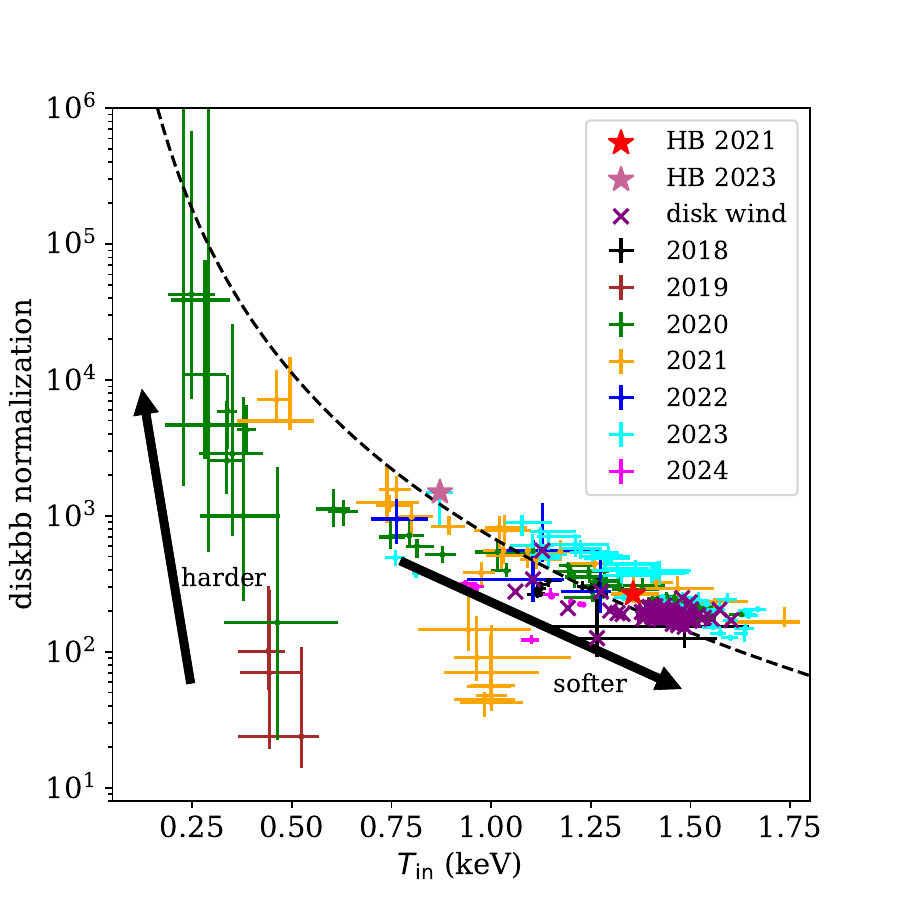}
    
    \caption{The fitting results showing the relationship between the inner-disk temperature and the disk normalization. The two arrows show the changing trend of hardness. The dashed line show constant $R_{\mathrm{in}}$ contours, calculated from the equation $norm_{\mathrm{diskbb}}=(R_{\mathrm{in}}/(f_{\mathrm{col}}^{2}D_{10}))^2\cos{i}$ \citep{Kubota1998_colorcorrection} and adopting a  temperature-dependent color correction factor $f_{\mathrm{col}}\propto T_{\mathrm{in}}^{1/4}$ \citep{Davis_Done_color_2006}.}
    \label{Tin_norm}
\end{figure}

We summarize the results of fitting the thermal component with \texttt{diskbb} for all observations in Fig.~\ref{Tin_norm}. Generally, the inner disk temperature is anti-correlated with the \texttt{diskbb} normalization factor. At the end of the 2020 observations and the 2019 observations, the inner disk temperature decreases to a quite low value at around 0.2~keV and the normalization factor of \texttt{diskbb} increases when the source returns from the soft state to the hard state (see the lower right points in the HID in Fig.~\ref{HID}). The estimates for these lower-temperature points ($T_{\rm{in}} \lesssim$ 0.4~keV) have large systematic uncertainties due to the exclusion of low-energy data and the low total counts in these spectra. When the source becomes brighter in the soft state, the \texttt{diskbb} normalization decreases with increasing temperature. For most of the time, the inner disk temperature is higher than 1~keV, indicating the dominance of the soft state of the source, which is also found in \cite{2015Capitanio_missinghard} through spectral-timing analysis of the 2006, 2008 and 2010 outbursts of 4U~1630--47.

It is worth noting that despite the similarity in the hardness, luminosity and timing properties between the two heartbeat states, we find they have quite different spectra and thermal features. We note that for both the 2021 and 2023 outbursts, sudden and large-amplitude changes in the inner disk temperature before, during and after the heartbeat state are observed (further details about the spectral and temperature changes can be found in Appendix~\ref{app:B}). In Section.~\ref{discussion}, we explore possible explanations for this large temperature shift.

\section{Phase-resolved fitting of the heartbeat state}\label{phase fitting}

In order to explore the physical changes during the quasi-periodic heartbeat oscillation, we follow the approach in \cite{Neilsen2011_physicsofheartbeat,Neilson2012_radiationpressure} to achieve a phase-folding to then determine phase bins with corresponding spectra we can analyze.  In this approach, we first pick out one typical cycle and calculate its cross correlation against the entire light curve. The maxima of the cross correlation function indicate starting points of new cycles. Based on \cite{Neilsen2011_physicsofheartbeat}, we choose the time with the highest count rate as the start of each period, corresponding to phase $\phi=0$. We then fold the entire light curve according to the starting points we have found, and iterate to achieve a new template profile each time by averaging the heartbeat cycles. This process of finding the starting points and calculating the averaged light curve in one period is iterated several times until we find an almost stable folded light curve.

\begin{figure*}
    \centering
    \includegraphics[width=0.4\linewidth]{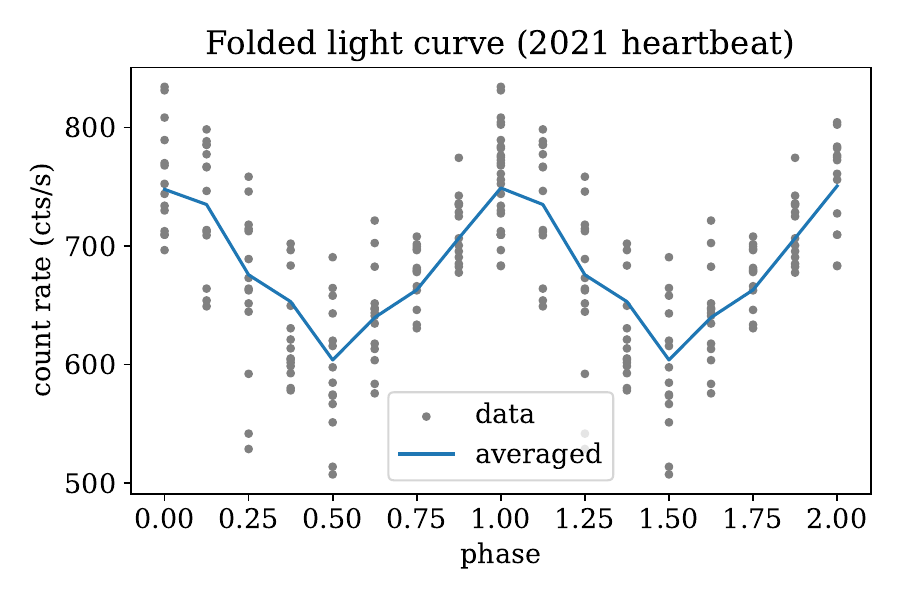}
    \includegraphics[width=0.4\linewidth]{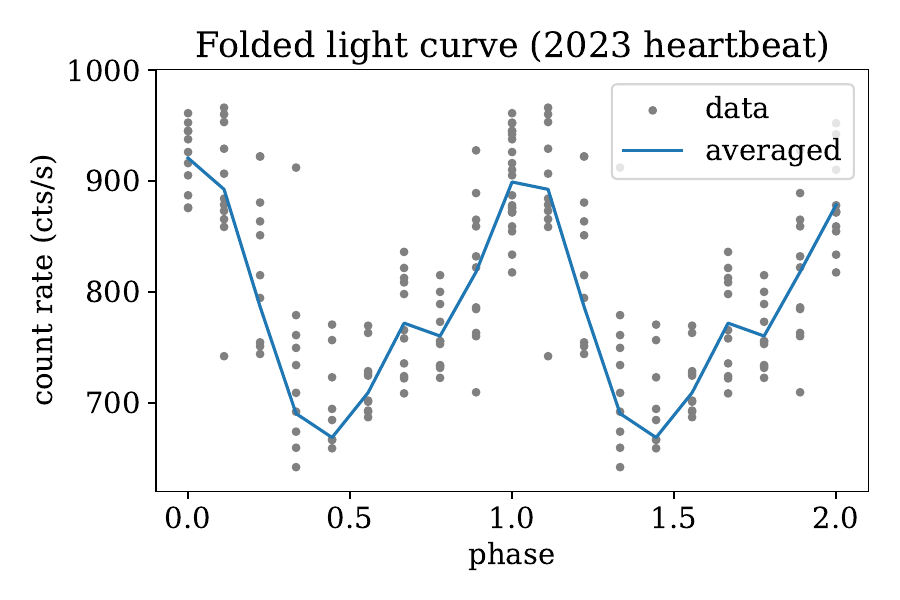}

    \caption{The phase-folded light curves for the 2021 heartbeat state (left) and 2023 heartbeat state (right). Gray points show the phase-mapped individual oscillations. The blue line shows the averaged folded light curve.}
    \label{phase_fold}
\end{figure*}

For each heartbeat observation, we employ a light curve of 300~s (Fig.~\ref{phase_lc}) to calculate the folded light curve. The oscillation period of the heartbeat in 2021 is about 18~s, and the period of the heartbeat in 2023 is about 20~s, which allows for more than 10 values for each phase and yields adequate statistics for phase-resolved fitting. The folded light curves are shown in Fig.~\ref{phase_fold}. Each phase bin corresponds to 2~s in an average period, so the folded light curve for the observation in 2021 has 9 bins, while the one for 2023 has 10 bins. Based on the folded light curves, we extract the data in different phases from different oscillation periods using the \texttt{extractor}\footnote{https://heasarc.gsfc.nasa.gov/lheasoft/ftools/headas/extractor.html} command in \texttt{heasoft} and combine all the data from the same phase together to derive the phase-resolved spectra. Each spectrum has more than 20000 counts on average. We then perform fits on each phase bin to obtain the spectral parameter evolution for different phases. In the following, we will present the phase-resolved fitting results with different models.

\subsection{Non-relativistic thermal model}
\begin{figure*}
    \centering
    \includegraphics[width=0.45\linewidth]{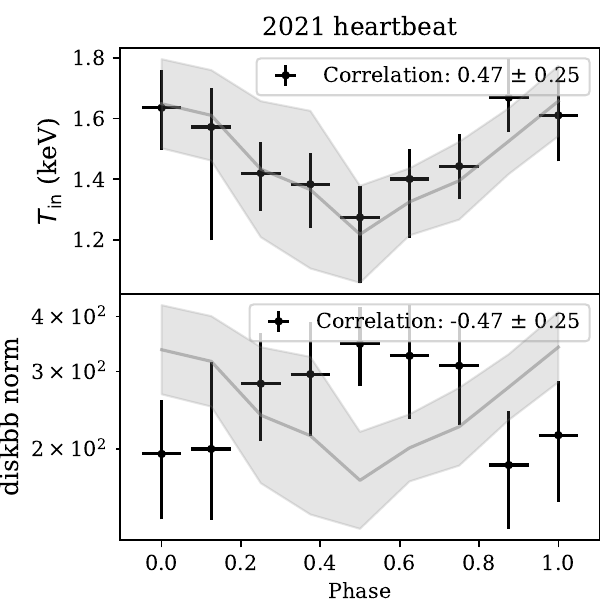}
    \includegraphics[width=0.45\linewidth]{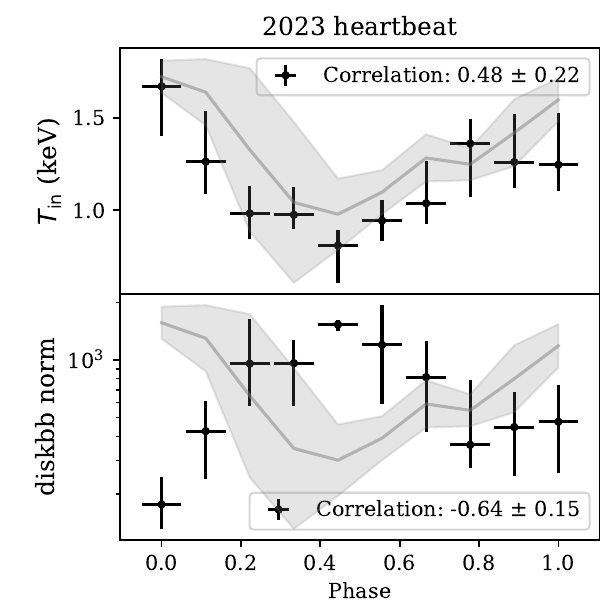} 
   
    \caption{The changes of disk parameters with phase in one heartbeat oscillation, fitted using the model \texttt{tbfeo $\times$ thcomp $\otimes$ diskbb} for the 2021 heartbeat state (left) and 2023 heartbeat state (right).The phase-folded lightcurve is overlaid in gray as a reference. The inner disk temperature ($T_\mathrm{in}$) is shown in the upper panels, and the disk normalization factor is shown in the lower panels. Pearson-correlation coefficients are shown for each, depicting the correlation between the phase-folded light curve profile and the spectral parameters.}
    \label{phase_diskbb}
\end{figure*}

In this part, we fit the phase-resolved spectra with the model: \texttt{tbfeo $\times$ thcomp $\otimes$ diskbb}. We calculate the correlation coefficient between the count rate and spectral parameter using the Pearson correlation coefficient. The error of the correlation coefficient is calculated at $1\sigma$ level. For both the heartbeat in 2021 and 2023, the count rate shows positive correlation with the inner disk temperature and negative correlation with the \texttt{diskbb} normalization factor, indicating that during the heartbeat oscillations, the higher the flux is, the smaller the inner disk radius is and the higher the inner disk temperature is. However, the coronal parameters are not observed to be correlated with the count rate at $1\sigma$ level in both heartbeats. How disk spectral parameters change with phase given by the fits is shown in Fig.~\ref{phase_diskbb}.

\subsection{Relativistic thermal component}

In order to see whether the fitting results are affected by the choice of the thermal component, we next replace \texttt{diskbb} with \texttt{kerrd}. \texttt{kerrd} describes the optically thick accretion disk around a Kerr black hole \citep{Ebisawa2003_kerrd}. Compared with the commonly used relativistic disk model for a Kerr black hole \texttt{kerrbb} \citep{Li2005_kerrbb}, the inner disk radius is also a free parameter in \texttt{kerrd}, thus enabling us to see the evolution of inner disk radius directly instead of estimating it from a non-relativistic thermal model or changes in spin. Meanwhile, the mass accretion rate parameter in \texttt{kerrd} allows us to see dynamical changes in the accretion process.

\begin{figure*}
    \centering
    \includegraphics[width=0.45\linewidth]{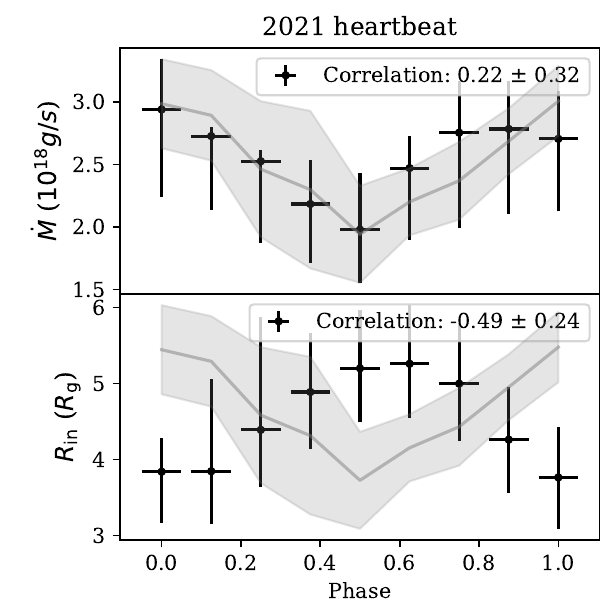}
    \includegraphics[width=0.45\linewidth]{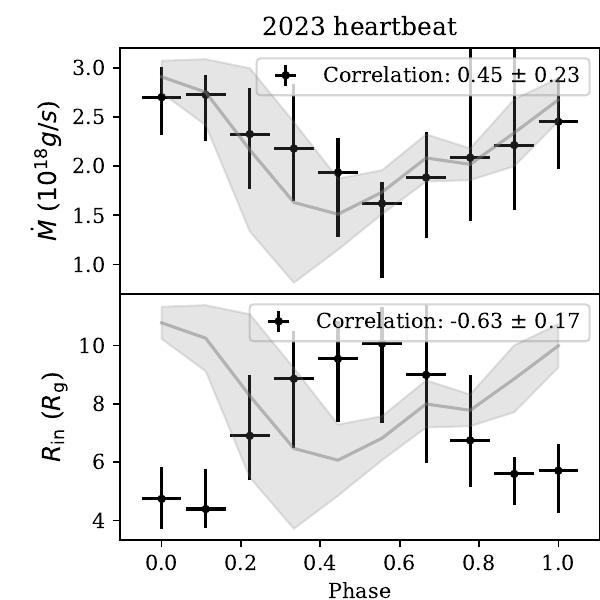}

    \caption{As in Fig.~\ref{phase_diskbb} but for fits with the spectral model \texttt{tbfeo $\times$ thcomp $\otimes$ kerrd}. Here, the mass accretion rate is shown in the upper panel and the inner disk radius ($R_\mathrm{in}$) is shown in the lower panel. The value of $R_\mathrm{in}$ in the unit of $R_\mathrm{g}$ is measured by setting and distance of the source at 10~kpc and its mass at $10~M_{\odot}$ in the fitting. However, the physical meaning of the value should be taken with caution because the source is located near the Galactic plane and has a high column density, therefore limiting a robust constraint on the distance and mass of the source \citep{Seifina2014_mass}.}
    \label{phase_kerrd}
\end{figure*}

With the model \texttt{tbfeo $\times$ thcomp $\otimes$ kerrd}, the changes of the coronal parameters are consistent with the previous ones despite the difference in the choice of the thermal model. At $\gtrsim 90\%$ confidence, we can see that the count rate is negatively correlated with the inner disk radius and positively correlated with the mass accretion rate, as shown in Fig.~\ref{phase_kerrd}.

\section{Timing analysis of the heartbeat state}\label{timing}
\subsection{Energy-resolved PDS}\label{energy-resolved pds}

We further explore the timing properties of the heartbeat state, first computing the PDS for photons in different energy bands. We fit each with a Lorentzian model, which is commonly used to fit PDS including those with QPOs \citep{Bachetti2022_fourier}. For the PDS in the heartbeat state, the variability can be modeled with three Lorentzian functions. We set a zero-centered Lorentzian for the broad band noise, one peaked at the heartbeat characteristic frequency, and a third to fit the noise at higher frequencies. From the central frequency and the full width half maximum (FWHM) of the Lorenzian function corresponding to the heartbeat peak, the quality factor of the 2021 heartbeat oscillation is $2.1^{+0.5}_{-0.9}$, and that of the 2023 heartbeat is $2.6^{+0.9}_{-1.6}$.

\begin{figure*}
    \centering
    \includegraphics[width=0.45\linewidth]{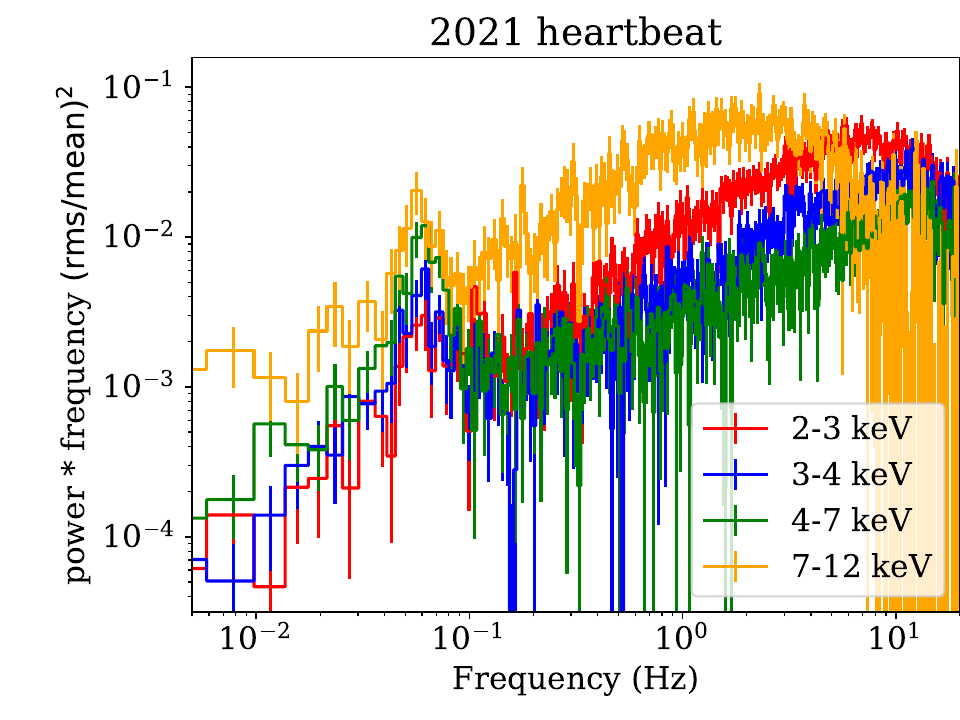}
    \includegraphics[width=0.45\linewidth]{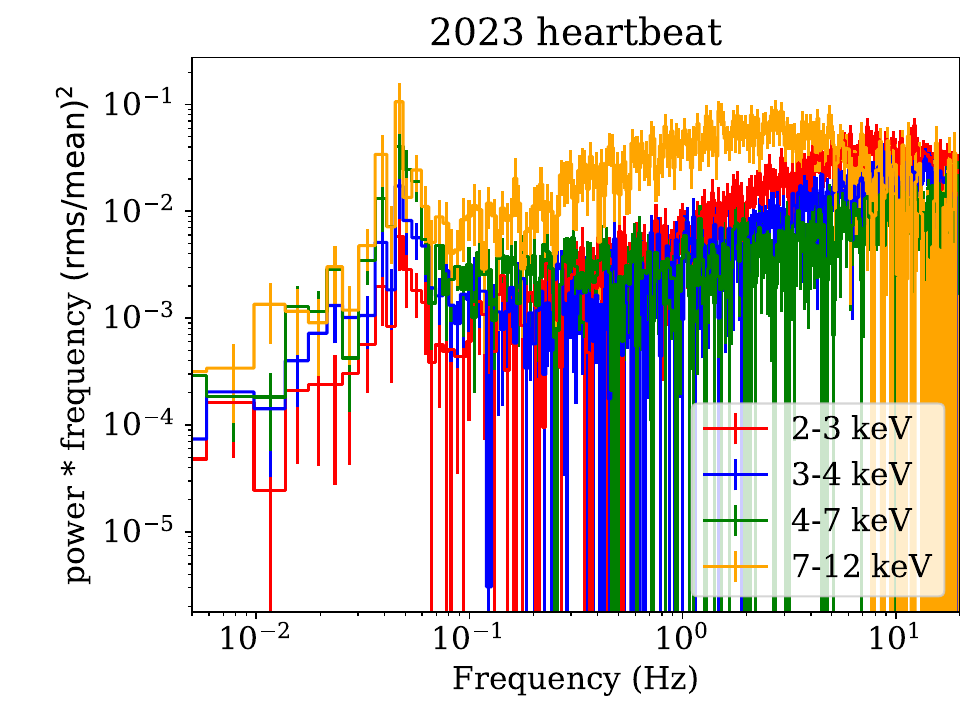}
    \includegraphics[width=0.5\linewidth]{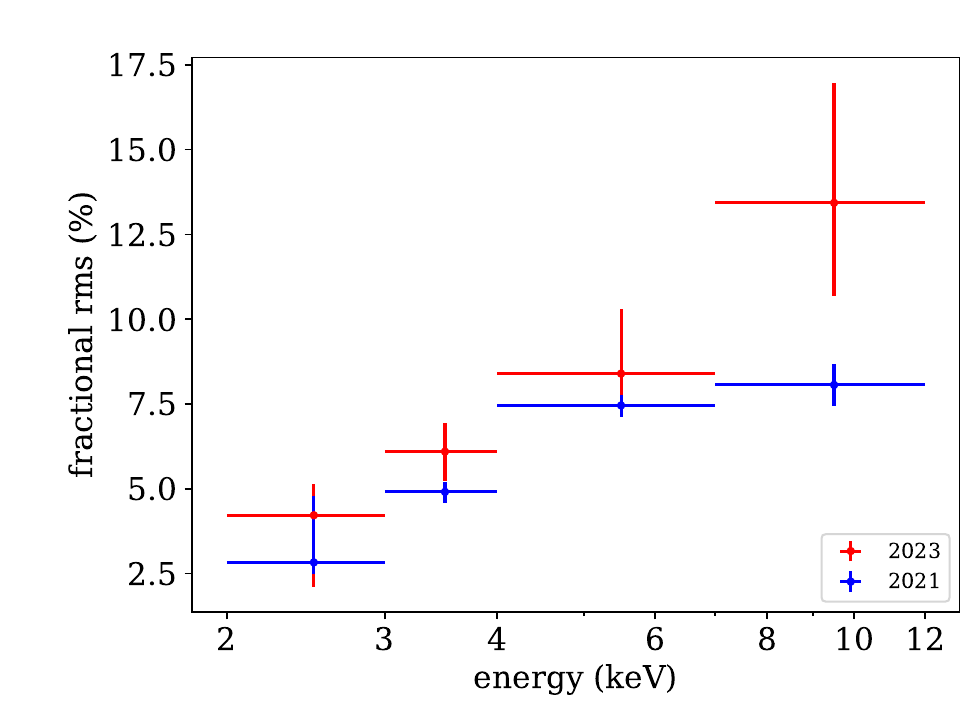}
    
    \caption{The energy dependence of the PDS properties of the heartbeat states in 2021 (left) and 2023 (right). Top panel: the energy-resolved PDS for 2--3~keV, 3--4~keV,4--7~keV and 7--12~keV. Lower panel: the energy dependence of the fractional rms in the heartbeat state.}
    \label{energy_pds}
\end{figure*}

The energy-resolved PDS and the corresponding energy-dependent fractional rms (here computed from the square root of the Lorentzian normalization) of the heartbeats are shown in Fig.~\ref{energy_pds}. The fractional rms corresponds to the amplitude of the heartbeat oscillation, and is found to be positively correlated with energy, while there is no clear correlation observed for the other two Lorentzian parameters.

\subsection{Energy-resolved time lag and coherence}\label{energy-resolved lag and coherence}
\begin{figure*}
    \centering
    \includegraphics[width=0.48\linewidth]{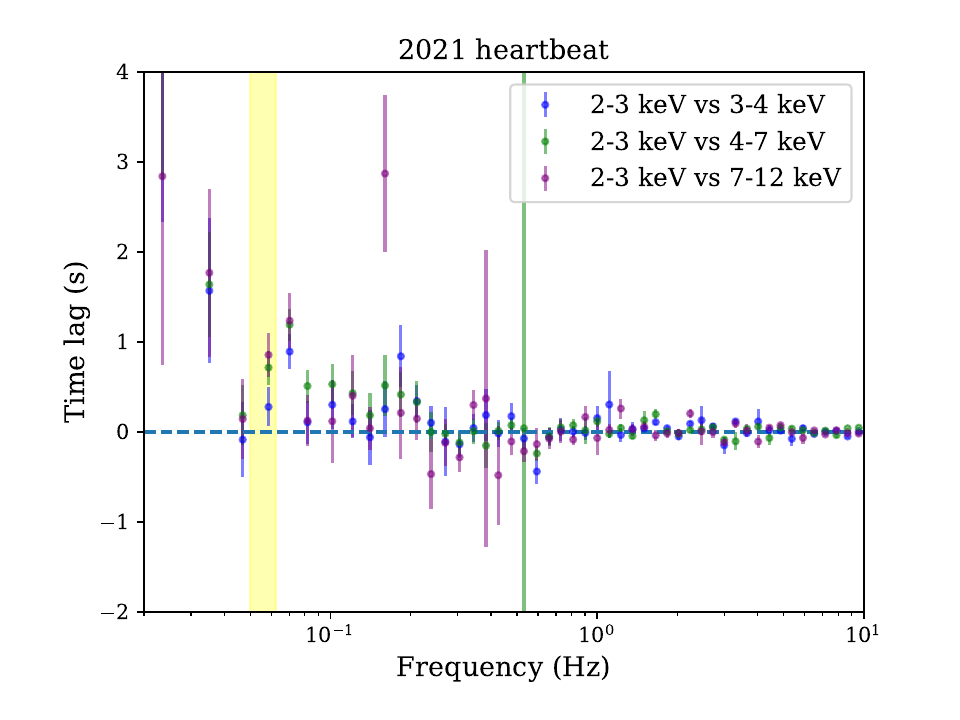}
    \includegraphics[width=0.48\linewidth]{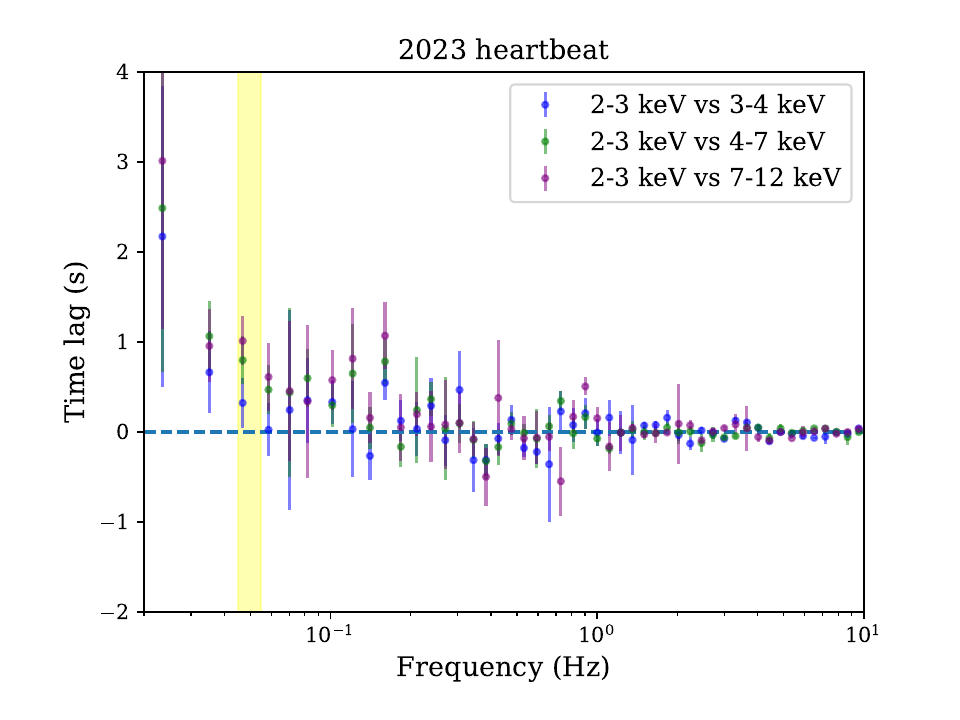}
    \includegraphics[width=0.48\linewidth]{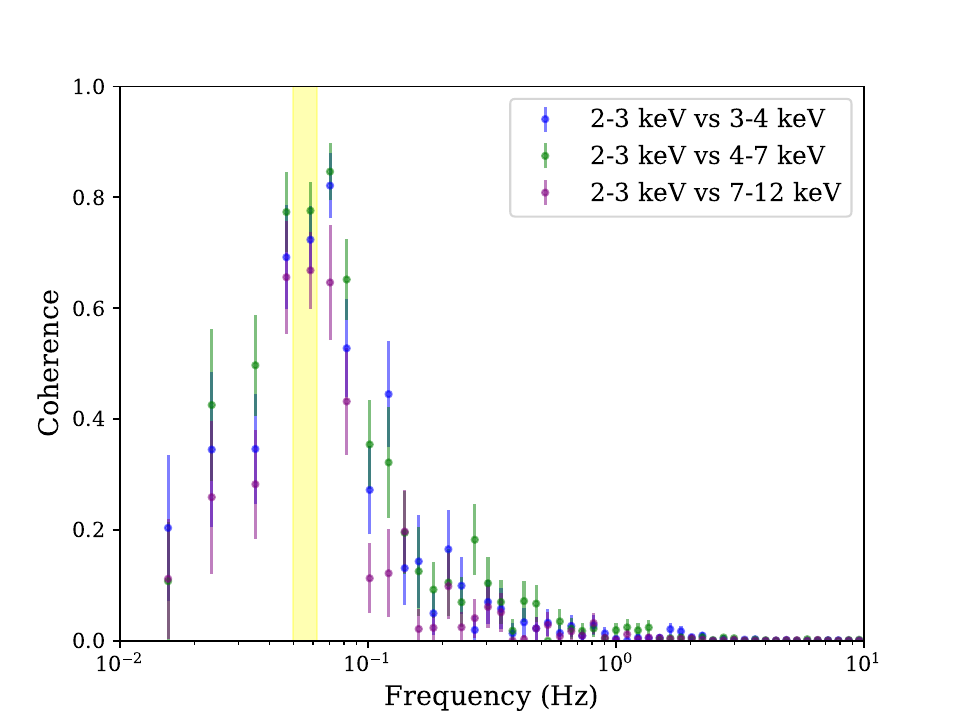}
    \includegraphics[width=0.48\linewidth]{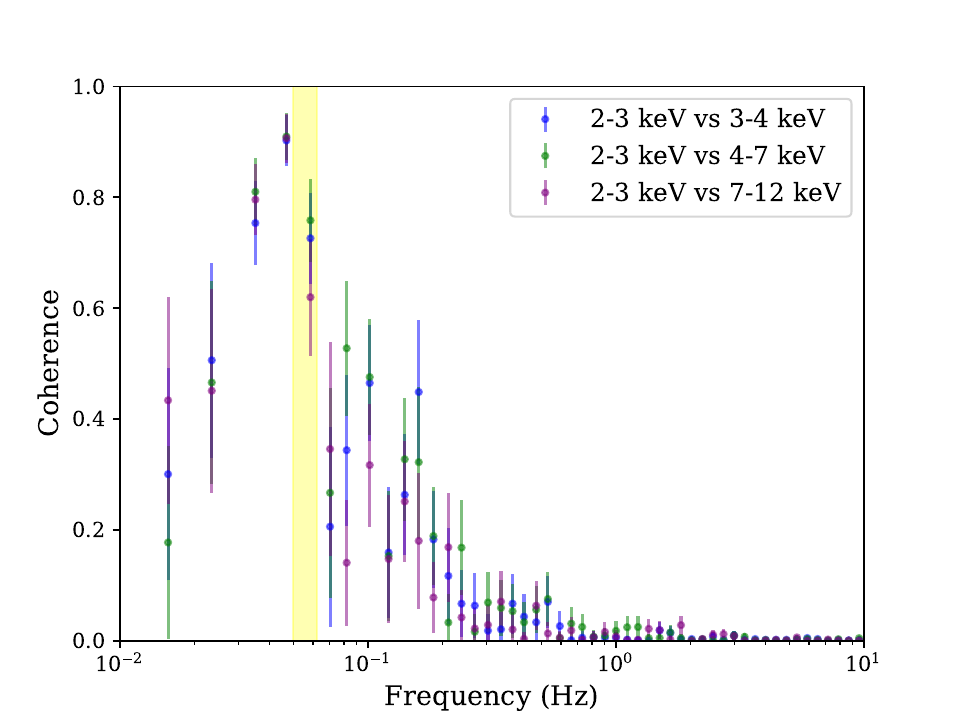}

    \caption{The energy dependence of lag (upper panels) and coherence (lower panels) of the heartbeat states in 2021 (left) and 2023 (right). The reference energy band is 2--3~keV, and the comparison bands are 3--4~keV, 4--7~keV, and 7--12~keV. The yellow region highlights the frequency range of  the heartbeat oscillations (from the PDS fits). Near the characteristic frequency of the heartbeats annotated by the yellow region, a hard lag of $\sim1~\mathrm{s}$ between 4--12~keV photons and 2--3~keV photons is observed.}
    \label{energy_lag_coherence}
\end{figure*}

Analysis of the lags between the photons of different energies is a powerful tool to study the origin of variability in data \citep{Jingyi_2021_diskcoronajetconnection,Neik2024_shortlag,Ma2021_highenergyQPO}, including heartbeats \citep{Yang2022_qrm2021}. In addition, coherence measures the degree of linear correlation between two time series \citep{Bachetti2022_fourier}, enabling us to assess the strength of the correlation between the oscillations of light curves of different energy bands. Here we choose the reference energy band to be 2--3~keV. We use \texttt{stingray} to calculate the frequency dependence of the lag and coherence of the chosen energy bands (3--4~keV, 4--7~keV, 7--12~keV) versus the reference energy band in the heartbeat states in 2021 and 2023, and the results are shown in Fig.~\ref{energy_lag_coherence}. From the coherence, we can see that the light curves of different energy bands are relatively highly linearly correlated (coherence $\gtrsim 0.8$) near the characteristic frequency of the heartbeats annotated by the yellow region. A hard lag of $\sim1$~s between 4--12~keV photons and 2--3~keV photons is observed near the heartbeat frequency. Although a few other frequency bins show evidence of similar-amplitude lags (e.g., at $\sim$ 3~mHz in the 2021 and 2023 heartbeat, at $\sim$ 0.105~Hz in the 2023 heartbeat), their coherence is relatively low. The non-linear relationship between signals of different energy bands makes the lag less straightforward to interpret, therefore we do not include these bins into our discussion.

\section{Discussion}\label{discussion}

\subsection{Stable inner disk radius in the soft state}

If we naively derive the inner disk radius from the \texttt{diskbb} normalization factor using the formula $norm_{\mathrm{diskbb}}=(R_{\mathrm{in}}/D_{10})^2\cos{i}$, it should be concluded from Fig.~\ref{Tin_norm} that $R_{\mathrm{in}}$ becomes smaller when the system becomes softer in the soft state. 

While at lower temperatures the disk emission can generally be well described by a multi-temperature disk blackbody spectrum, non-thermal effects such as the Compton scattering in the disk atmosphere may produce deviations from the multi-temperature disk blackbody spectrum, especially when the inner disk temperature is high. This is normally taken into account by introducing a color correction factor $f_{\mathrm{col}}$ (also called spectral hardening factor) \citep{color_correction,Kubota1998_colorcorrection}. Moreover, it has been found in previous studies that the color correction factor has strong dependence on the disk temperature \citep{Kubota_Makishima_color_2004,Davis_Done_color_2006}, which can be approximately described by $f_{\mathrm{col}}\propto T_{\mathrm{in}}^{1/4}$. Based on this scaling relation and the equation $norm_{\mathrm{diskbb}}=(R_{\mathrm{in}}/(f_{\mathrm{col}}^{2}D_{10}))^2\cos{i}$ \citep{Kubota1998_colorcorrection}, we overlay the constant $R_{\mathrm{in}}$ contour in Fig.~\ref{Tin_norm}, showing that $R_{\mathrm{in}}$ is essentially constant when the system is in the soft state. The $R_{\rm{in}}$ of the brown points that belong to the 2019 outburst in the bottom-left corner of Fig.~\ref{Tin_norm} looks slightly lower than the $R_{\rm{in}}$ in the soft state in other years of outbursts. Similarly, a slightly lower value of $R_{\rm{in}}$ at the beginning of the 2021 outburst is also found from reflection modeling (see the first two points in Fig.~\ref{reflection fit}). A smaller $R_{\rm{in}}$ compared to the ISCO is not consistent with the standard picture. However, there are a few uncertainties that may be responsible for such a result: $i)$ \texttt{diskbb} is a Newtonian model and can only provide a very rough estimate of the location of the inner edge of the accretion disk, $ii)$ the faintness of the source and the limited quality of these spectra.

Similar results on the inner disk radius have also been found in Swift~J1727--1613 (K\"{o}nig et al. 2025, in preparation).  We note a long-history of evidence from the thermal disk emission for soft states exhibiting a stable accretion-disk inner radius (e.g., detailed analysis of several decades of observations of LMC~X--3; see \cite{Jack2010_LMCX3}, and references therein). 

\subsection{Heartbeat mechanism}

Inner disk radiation instability is a popular candidate mechanism for producing the heartbeat oscillation \citep{Lightman1974_instability,Done2007review,Neilson2012_radiationpressure}. In that scenario, the dominance of the radiation pressure in the inner disk region and the lack of balancing mechanism give rise to a thermal-viscous instability cycle resulting in the heartbeat oscillations in the light curve \citep{Done2007review,Neilsen2011_physicsofheartbeat}. 

To gain a better understanding of different physics processes behind this cycle behavior, it is useful to refer to the ``S-curve'' showing the relationship between the mass and temperature at a certain disk radius \citep[e.g., see Fig.~7 in][]{Done2007review}. The curve has three branches: a lower viscous slow rise (where the heating is balanced by radiative cooling and the temperature increases slowly with mass), an upper slow viscous fall (where the heating is balanced by advective cooling and the temperature decreases slowly with mass) and the middle thermal viscous cycle (where the accretion process switches quickly between the two stable branches) \citep{Wang2024_IGRJ17091}. This cycled oscillation causes a density wave to propagate along the disk and results in the fast changes in the mass accretion rate, inner disk radius, and temperature. The fast variability of disk parameters in our phase-resolved fitting results (Sect.~\ref{phase fitting}) is consistent with the scenario given by the radiation pressure instability theory. Previous phase-resolved and time-resolved spectral fitting of other sources reached a similar conclusion \citep{Belloni1997_time,Neilsen2011_physicsofheartbeat,Rawat2022_GRS1915_timeresolved}.

Besides the phase-resolved fitting of the heartbeat spectra, the energy dependence of the PDS properties and the lags between the photons of different energy bands can also be informative in revealing the origin of the heartbeat. We observe a hard lag near the heartbeat characteristic frequency of about one second (similar hard lag at $\sim$ mHz has also been observed in other sources, e.g., GX 339--4 \citep{Uttley2011_hardlag_GX339} and MAXI~J1820+070 \citep{2024Uttley_Malzac}). This timescale is several orders of magnitude longer than those expected for reverberation. Instead, a possible explanation for such a large lag is the viscous propagation of mass accretion fluctuations in the disk \citep{Uttley2011_hardlag_GX339,Lyubarskii1997_accretion_noise}. The broad peaks of the coherence plots may indicate that in a broader frequency range near the heartbeat frequency, photons of different energies are causally related. In the heartbeat state, with the disappearance and refilling of the inner edge of the disk, the characteristic frequency of the viscous propagation lag may slightly change during this process. The frequency range of the coherence peak, corresponding to a time difference of several seconds, is comparable to the viscous time differences for the changing inner disk radius \citep{Belloni1997_time}.

Moreover, a positive correlation between fractional rms of the heartbeat variability and energy is observed. This trend is also observed by \citet{Yang2022_qrm2021} to reach an energy of 100~keV. The authors conclude that their \textit{Insight}-HXMT data constrain the origin of the heartbeat to the corona region. However, our phase-resolved fitting results do not show strong correlation between the count rate and the coronal parameters, which is also the case in the time-resolved fitting of the heartbeat state of GRS~1915+105 in \cite{Rawat2022_GRS1915_timeresolved}. Instead, we suggest that disk photons have a temperature modulation, seeding variability which is magnified by the coronal Compton scattering.  This could plausibly account for the positive correlation between the fractional rms and energy. A similar scenario has also been proposed in \cite{2024Uttley_Malzac} and supported by analytical calculations.

From the phase-resolved fitting results combined with the radiation pressure instability scenario, we can phenomenologically explain the changes of the inner disk temperature before, during, and after the heartbeat. Before the thermal viscous cycle of the heartbeat begins, the disk accretion process should be on one of the stable branches and the temperature at each radius is locally stable; at the heartbeat cycles onset, the temperature is seen to oscillate  over the heartbeat pattern.  Because of this, the aggregate spectrum of the heartbeat state blends a range of different thermal components (i.e., a range of disk temperatures). This results in a marked change in the temperature obtained for fitting the time-averaged observation compared to the adjacent data. This is apparent from both the 2021 and 2023 heartbeat states (see Appendix~\ref{app:B}).

In the above discussion, different mechanisms are combined into a more complete scenario of the heartbeat: at a certain strength of mass accretion, the inner disk radiation pressure instability triggers the heartbeat-like oscillations of disk properties. This variation then propagates from the inner part to the outer part of the disk, and extends to higher energy photons when disk photons are scattered by the corona. This scenario highlights the close connection among different parts of the disk-corona system in contributing to the spectral and timing variability seen in the data. Similar discussions about the disk-corona interaction during the heartbeat state of GRS~1915+015 can be found in \cite{yan2018corona_disc_1915}, and a more general review about that in different systems can be found in \cite{Swank2001}.

The precise origin of the different timescales and spectral properties seen between different heartbeat states in the same source is an interesting question but outside of the scope of our work. See \cite{Belloni1997_time} for a nice discussion about the relationship between the heartbeat period and the radius of the disappearing region in the disk.

\subsection{The production of wind}

A number of BHXRBs have been found to show absorption lines in their spectra which are associated with wind features. Examples include GRS~1915+015 \citep{Zoghbi2016_diskwindconnection1915,Miller2016_accretionwind1915,Liu2022_1915wind}, IGR~J17091–-3624 \citep{Wang2024_IGRJ17091}, MAXI~J1803--298 \citep{Zhang2024_1803wind}, as well as the subject of this paper, 4U~1630--47 \citep{Pahari2018_spin_wind,Trueba2019_4U1630wind}. There are mainly three candidates to explain the driving mechanism for the wind: radiation-, thermal-, or magnetic-driving mechanism \citep[see Sect.~10 in][and references therein, for a general review and more detailed explanation for different mechanisms]{Done2007review}.

Our spectral analysis shows that the wind absorption features appear in certain spectral states: the SIMS and soft state (see the HID and HRD in Fig.~\ref{HID} and Fig.~\ref{frms-hardness}). In addition, Fig.~\ref{Tin_norm} shows that the temperature of the observations with wind features is concentrated near a quite high value of about 1.5~keV, indicating the close connection between the production of wind and the strength and temperature of the thermal component. The luminosity of the source when wind features appear is well-below the peak brightness and is therefore unlikely to be super-Eddington \citep[also found in][]{Trueba2019_4U1630wind,Pahari2018_spin_wind}. This indicates that the radiation pressure may not be high enough to drive the wind. In addition to the thermal-driving scenario, the magnetic-driving mechanism may also play an important role in the appearance of winds in 4U~1630--47, as is discussed in detail in \cite{Miller_2015,Trueba2019_4U1630wind}.

Based on the averaged wind parameters from the \texttt{XSTAR} fits, the outflowing velocity of the wind is $\sim 0.003~c$. Using the spherical approximation, the launching radius of the wind is $\sim 10^{10}~\rm{cm}$, corresponding to $\sim 10^{3-4}~R_{\rm{g}}$ for a $10~M_{\odot}$ black hole. The mass loss rate by the wind is $\sim 10^{18}~\rm{g/s}$, corresponding to $\sim 10\%$ Eddington accretion rate for a $10~M_{\odot}$ black hole. Therefore, the mass loss rate by the wind is comparable to the mass accretion rate of the source. These estimations of the wind properties are consistent with other studies of the source \citep{Kubota2007} and comparable to studies of other sources in order of magnitude, for example, GRS~1915+015 \citep{Zoghbi2016_diskwindconnection1915} and MAXI~J1803--298 \citep{Zhang2024_1803wind}. The values of the launching radius and velocity of the wind further support the thermal/magnetic origin of the wind, instead of being radiation-driven.

In our study, no strong correlation between the appearance of the wind and the heartbeat is observed. For example, in the 2018 and 2022 outbursts, the wind is observed but not the heartbeat, and conversely for the 2021 outburst. In \cite{Hori_2018_wind_evolution,2014Trigo_wind_disappear}, the ionization degree of the wind of this source is observed to be positively correlated with the luminosity, and spectral wind absorption features disappear when the luminosity is extremely high and the wind is over-ionized. In our case, the heartbeat luminosity is lower than the luminosity of wind observations, therefore over-ionization may not be a plausible explanation for the absence of wind features during the heartbeats. However, in other BHXRB systems, notably GRS 1915+015 \citep{Zoghbi2016_diskwindconnection1915} and IGR~J17091--3624 \citep{Wang2024_IGRJ17091}, the heartbeat state is observed simultaneously with the wind, making possible a link between the instability giving rise to the heartbeat and wind production, at least for those systems.

\section{Conclusions}\label{conclusion}

In this paper, we present spectral and timing analysis of the \textit{NICER} observations of the source 4U~1630--47 from 2018 to 2024. Our main results include: 

\begin{enumerate}
    \item We fit all the spectra with the disk-corona model \texttt{tbfeo $\times$ thcomp $\otimes$ diskbb} and find some spectra showing broadened iron line reflection features in the HIMS and SIMS, and some showing disk wind absorption features in the SIMS and soft state. We fit the reflection spectra with \texttt{relxillCp} and find a stable and untruncated disk in the intermediate states; For the disk wind absorption features, we fit them with \texttt{XSTAR} and find a stable, highly ionized wind with high column density across different outbursts.
    \item Through the analysis of the light curves and power density spectra, we find two observations in 2021 and 2023 with heartbeat features. The light curves of these observations oscillate quasi-periodically at a frequency around 0.05\,Hz and have a heartbeat-like profile. Through phase-resolved fitting of the spectra, we find the flux to be correlated with disk parameters: the inner disk temperature, inner disk radius and mass accretion rate, while no strong correlation with coronal parameters is found. This fast disk oscillation supports the inner disk radiation pressure instability as the driving mechanism of the heartbeats. Moreover, a hard lag on the time scale of a second is observed near the characteristic frequency of the heartbeat, which can be explained by the viscous propagation of mass accretion fluctuations in the disk. Combining the phase-resolved spectral analysis and lag analysis, the scenario where the disk-originated oscillation is then magnified by the corona scattering is a possible explanation for the positive relationship between the fractional rms and energy.
    \item  The observations with disk wind features are found to have a high disk temperature, while their luminosities are well below the peak brightness (so unlikely to be super-Eddington).  Accordingly, the wind is more likely to be thermally driven or magnetically driven instead of radiatively driven.
\end{enumerate}

\vspace{0.5cm}
{\bf Acknowledgments --}
The authors would like to thank the anonymous referee for valuable suggestions that improved the manuscript. N.F. thanks Honghui Liu, Zuobin Zhang, Yujia Song, Josephine Wong and Niek Bollemeijer for useful discussions. 
The work of N.F. and C.B. was supported by the National Natural Science Foundation of China (NSFC), Grant No.~11973019, 12250610185, and 12261131497, and the Natural Science Foundation of Shanghai, Grant No. 22ZR1403400. J.F.S and O.K. gratefully acknowledge support from NICER Guest Observer grants 80NSSC23K1096 and 80NSSC24K0535. Y.Z. acknowledges support from the Dutch Research Council (NWO) Rubicon Fellowship, file no. 019.231EN.021. O.K. acknowledges NICER GO funding 80NSSC23K1660.

\vspace{1cm}
\appendix
\vspace*{-10pt}
\section{Quantitative definition of winds}\label{app:A}
\begin{figure*}[htbp]
    \centering
    \includegraphics[width=0.98\linewidth]{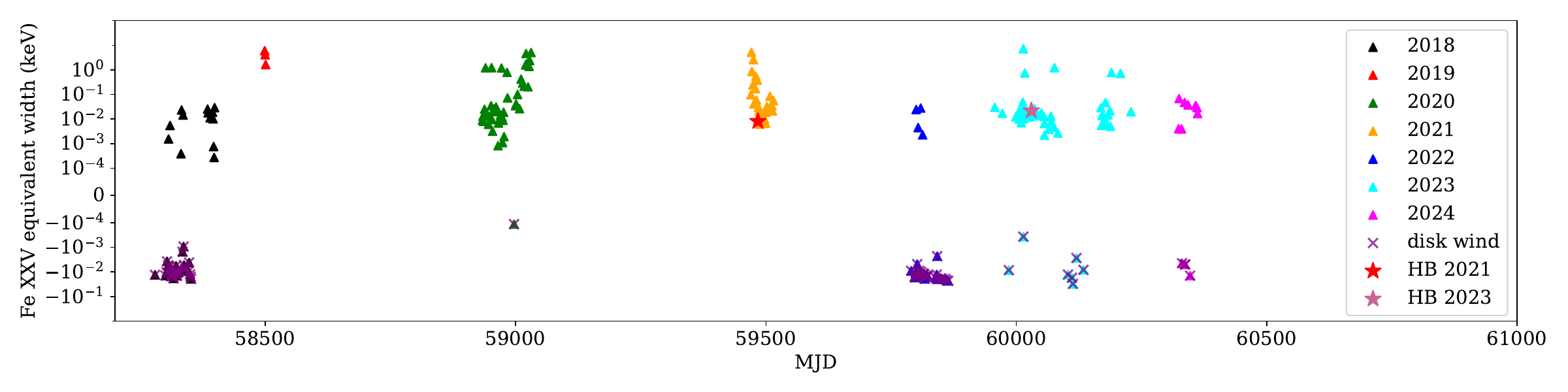}
    \includegraphics[width=0.98\linewidth]{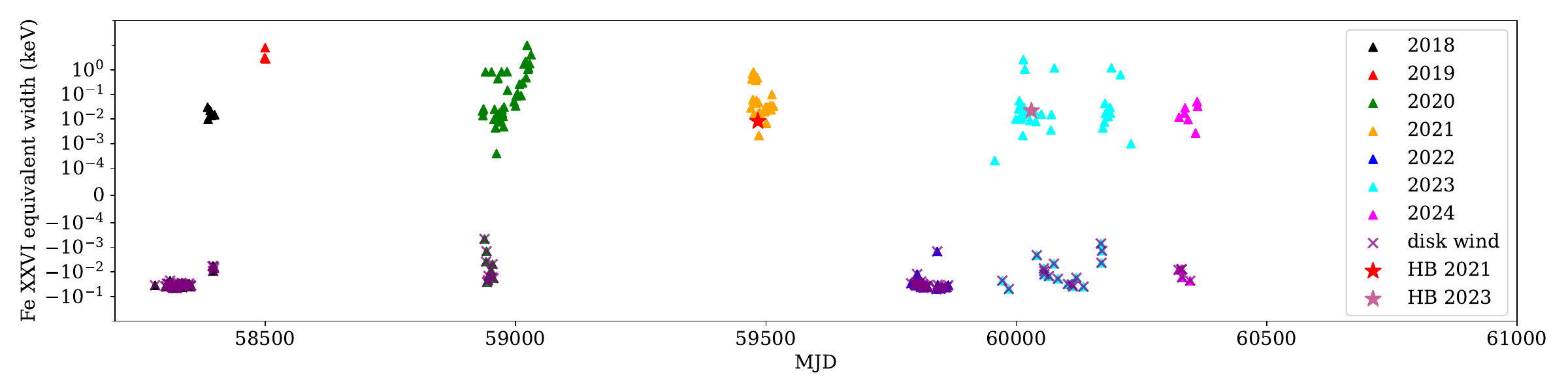}
    
    \caption{The upper bound of the equivalent width of the the Fe XXV (upper panel) and Fe XXVI (lower panel) absorption line. The purple crosses show the spectra we identify with wind features from a below zero upper bound of either of the two lines.}
    \label{gaussian_ew}
\end{figure*}

Besides judging from the shape of the fitting residuals, we identify spectra with wind absorption features quantitatively. We add two Gaussian functions to the continuum model \texttt{tbfeo $\times$ thcomp $\otimes$ diskbb}, with their centers fixed at the energies of the Fe XXV and Fe XXVI lines respectively. We calculate the equivalent width of the two absorption lines and show the upper bound of equivalent width (at a 90\% confidence level) in Fig.~\ref{gaussian_ew}. If the upper limit of the equivalent width for either of the absorption lines in a spectrum falls below zero, we identify it as exhibiting wind absorption. This quantitative criterion is consistent with our qualitative assessment.

\section{Heartbeat spectra}\label{app:B}

\begin{figure*}[htbp]
    \centering
    \includegraphics[width=0.45\linewidth]{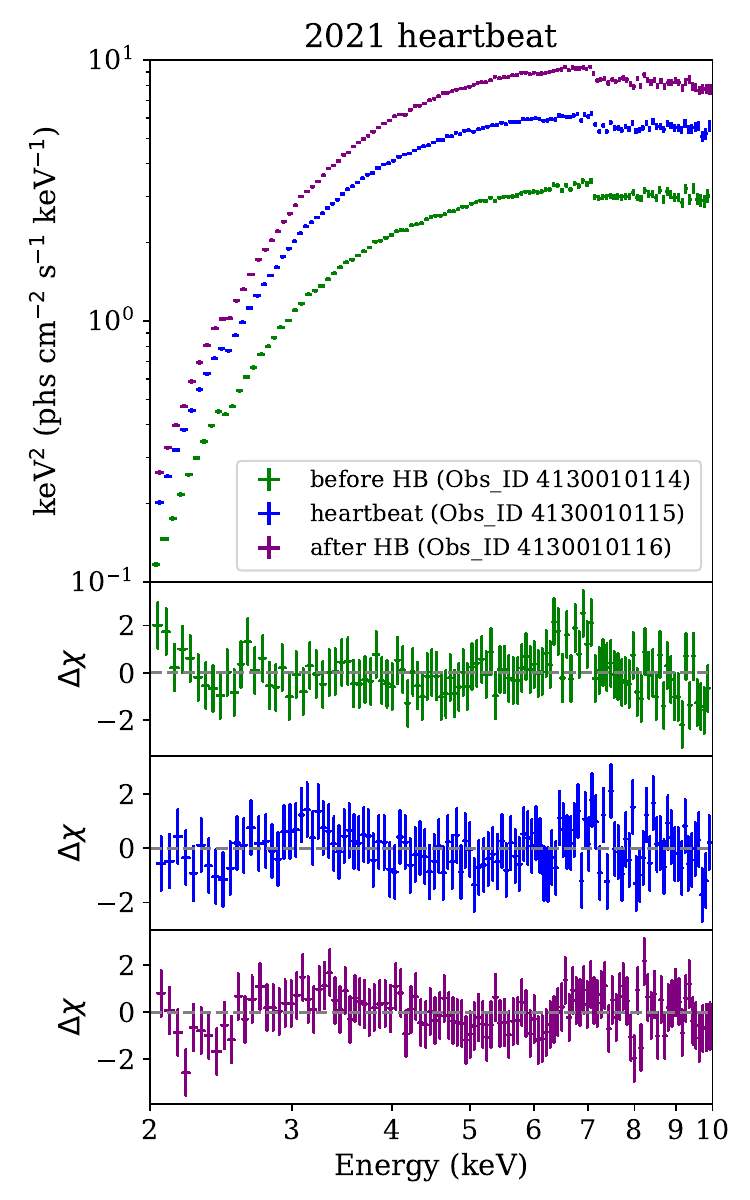}
    \includegraphics[width=0.45\linewidth]{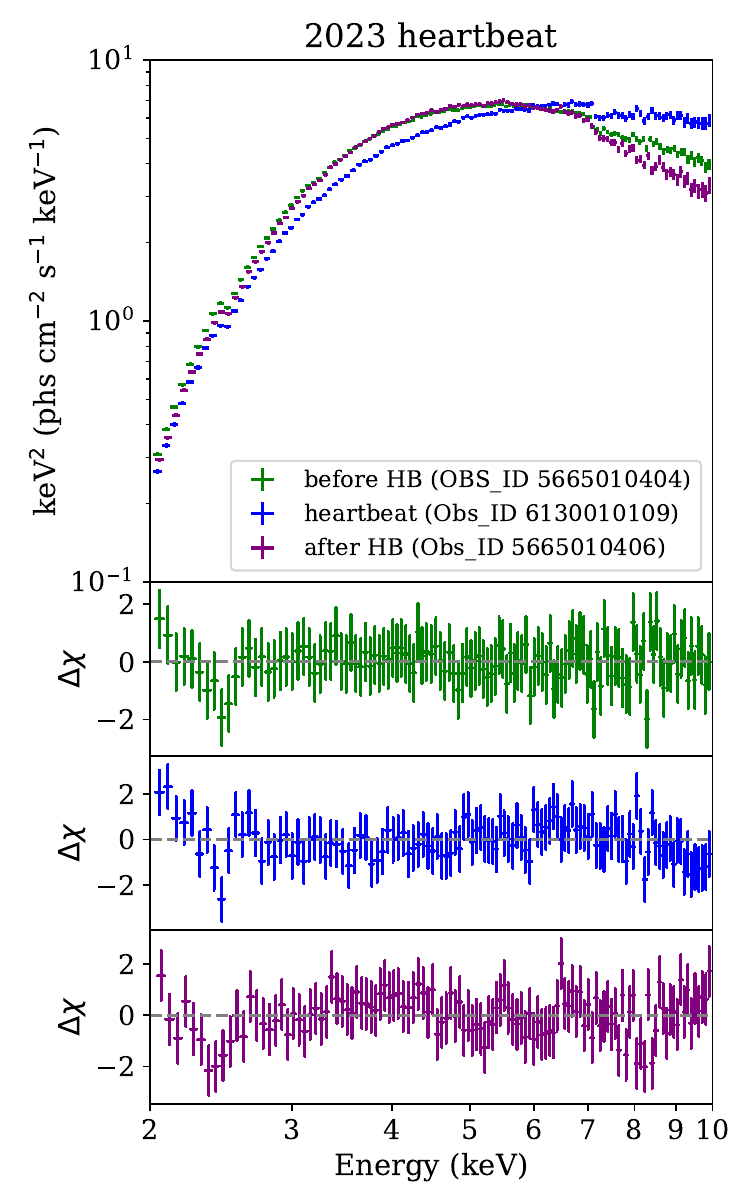}
    
    \caption{The upper panel shows the the spectra of the heartbeat state in 2021 (left panel) and 2023 (right panel), along with spectra from observations taken before and after the heartbeat state. The lower panels show the residuals of the fits of these spectra with the model \texttt{tbfeo $\times$ thcomp $\otimes$ diskbb}.}
    \label{hb_spec}
\end{figure*}

In Fig.~\ref{hb_spec}, we present the spectra of the heartbeat state, along with spectra from observations taken before and after this state. The heartbeat spectra differ significantly from those of the nearby observations, with notable changes in the inner disk temperature ($T_{\rm{in}}$). In 2021, $T_{\rm{in}}$ shifts from $\sim$ 0.75~keV before the heartbeat, to $\sim$ 1.35~keV during the heartbeat, and then to $\sim$ 1.75~keV afterward. In 2023, $T_{\rm{in}}$ changes from $\sim$ 1.2~keV before the heartbeat, to $\sim$ 0.8~keV during it, and then to $\sim$ 1.6~keV afterward.

\bibliographystyle{aasjournal}
\bibliography{ref}

\begin{thebibliography}{}
\expandafter\ifx\csname natexlab\endcsname\relax\def\natexlab#1{#1}\fi
\providecommand{\url}[1]{\href{#1}{#1}}
\providecommand{\dodoi}[1]{doi:~\href{http://doi.org/#1}{\nolinkurl{#1}}}
\providecommand{\doeprint}[1]{\href{http://ascl.net/#1}{\nolinkurl{http://ascl.net/#1}}}
\providecommand{\doarXiv}[1]{\href{https://arxiv.org/abs/#1}{\nolinkurl{https://arxiv.org/abs/#1}}}

\bibitem[{{Alabarta} {et~al.}(2020){Alabarta}, {Altamirano}, {M{\'e}ndez},
  {C{\'u}neo}, {Zhang}, {Remillard}, {Castro}, {Ludlam}, {Steiner}, {Enoto},
  {Homan}, {Arzoumanian}, {Bult}, {Gendreau}, {Markwardt}, {Strohmayer},
  {Uttley}, {Tombesi}, \& {Buisson}}]{Alabarta2020_1727HRD}
{Alabarta}, K., {Altamirano}, D., {M{\'e}ndez}, M., {et~al.} 2020, \mnras, 497,
  3896, \dodoi{10.1093/mnras/staa2168}

\bibitem[{{Arnaud}(1996)}]{Xspec_Arnaud}
{Arnaud}, K.~A. 1996, in Astronomical Society of the Pacific Conference Series,
  Vol. 101, Astronomical Data Analysis Software and Systems V, ed. G.~H.
  {Jacoby} \& J.~{Barnes}, 17

\bibitem[{{Bachetti} \& {Huppenkothen}(2022)}]{Bachetti2022_fourier}
{Bachetti}, M., \& {Huppenkothen}, D. 2022, arXiv e-prints, arXiv:2209.07954,
  \dodoi{10.48550/arXiv.2209.07954}

\bibitem[{{Belloni} {et~al.}(1997){Belloni}, {M{\'e}ndez}, {King}, {van der
  Klis}, \& {van Paradijs}}]{Belloni1997_time}
{Belloni}, T., {M{\'e}ndez}, M., {King}, A.~R., {van der Klis}, M., \& {van
  Paradijs}, J. 1997, \apjl, 488, L109, \dodoi{10.1086/310944}

\bibitem[{{Belloni}(2010)}]{Belloni2010}
{Belloni}, T.~M. 2010, in Lecture Notes in Physics, Berlin Springer Verlag, ed.
  T.~{Belloni}, Vol. 794, 53, \dodoi{10.1007/978-3-540-76937-8_3}

\bibitem[{{Belloni} \& {Motta}(2016)}]{Belloni2016}
{Belloni}, T.~M., \& {Motta}, S.~E. 2016, in Astrophysics and Space Science
  Library, Vol. 440, Astrophysics of Black Holes: From Fundamental Aspects to
  Latest Developments, ed. C.~{Bambi}, 61, \dodoi{10.1007/978-3-662-52859-4_2}

\bibitem[{{Belloni} {et~al.}(2011){Belloni}, {Motta}, \&
  {Mu{\~n}oz-Darias}}]{Belloni_Motta_2011}
{Belloni}, T.~M., {Motta}, S.~E., \& {Mu{\~n}oz-Darias}, T. 2011, Bulletin of
  the Astronomical Society of India, 39, 409, \dodoi{10.48550/arXiv.1109.3388}

\bibitem[{{Bollemeijer} {et~al.}(2024){Bollemeijer}, {Uttley}, {Basak},
  {Ingram}, {van den Eijnden}, {Alabarta}, {Altamirano}, {Arzoumanian},
  {Buisson}, {Fabian}, {Ferrara}, {Gendreau}, {Homan}, {Kara}, {Markwardt},
  {Remillard}, {Sanna}, {Steiner}, {Tombesi}, {Wang}, {Wang}, \&
  {Zoghbi}}]{Neik2024_shortlag}
{Bollemeijer}, N., {Uttley}, P., {Basak}, A., {et~al.} 2024, \mnras, 528, 558,
  \dodoi{10.1093/mnras/stad3912}

\bibitem[{{Buisson} {et~al.}(2019){Buisson}, {Fabian}, {Barret}, {F{\"u}rst},
  {Gandhi}, {Garc{\'\i}a}, {Kara}, {Madsen}, {Miller}, {Parker}, {Shaw},
  {Tomsick}, \& {Walton}}]{Buisson2019}
{Buisson}, D.~J.~K., {Fabian}, A.~C., {Barret}, D., {et~al.} 2019, \mnras, 490,
  1350, \dodoi{10.1093/mnras/stz2681}

\bibitem[{{Capitanio} {et~al.}(2015){Capitanio}, {Campana}, {De Cesare}, \&
  {Ferrigno}}]{2015Capitanio_missinghard}
{Capitanio}, F., {Campana}, R., {De Cesare}, G., \& {Ferrigno}, C. 2015,
  \mnras, 450, 3840, \dodoi{10.1093/mnras/stv687}

\bibitem[{{Casella} {et~al.}(2005){Casella}, {Belloni}, \&
  {Stella}}]{Casella2005_ABC}
{Casella}, P., {Belloni}, T., \& {Stella}, L. 2005, \apj, 629, 403,
  \dodoi{10.1086/431174}

\bibitem[{{Choudhury} {et~al.}(2015){Choudhury}, {Bhatt}, \&
  {Bhattacharyya}}]{Choudhury2015}
{Choudhury}, M., {Bhatt}, N., \& {Bhattacharyya}, S. 2015, \mnras, 447, 3960,
  \dodoi{10.1093/mnras/stu2742}

\bibitem[{{Davis} {et~al.}(2006){Davis}, {Done}, \&
  {Blaes}}]{Davis_Done_color_2006}
{Davis}, S.~W., {Done}, C., \& {Blaes}, O.~M. 2006, \apj, 647, 525,
  \dodoi{10.1086/505386}

\bibitem[{{De Marco} {et~al.}(2021){De Marco}, {Zdziarski}, {Ponti},
  {Migliori}, {Belloni}, {Segovia Otero}, {Dzie{\l}ak}, \&
  {Lai}}]{2021DeMarco_1820}
{De Marco}, B., {Zdziarski}, A.~A., {Ponti}, G., {et~al.} 2021, \aap, 654, A14,
  \dodoi{10.1051/0004-6361/202140567}

\bibitem[{{D{\'\i}az Trigo} {et~al.}(2014){D{\'\i}az Trigo}, {Migliari},
  {Miller-Jones}, \& {Guainazzi}}]{2014Trigo_wind_disappear}
{D{\'\i}az Trigo}, M., {Migliari}, S., {Miller-Jones}, J.~C.~A., \&
  {Guainazzi}, M. 2014, \aap, 571, A76, \dodoi{10.1051/0004-6361/201424554}

\bibitem[{{Done} {et~al.}(2007){Done}, {Gierli{\'n}ski}, \&
  {Kubota}}]{Done2007review}
{Done}, C., {Gierli{\'n}ski}, M., \& {Kubota}, A. 2007, \aapr, 15, 1,
  \dodoi{10.1007/s00159-007-0006-1}

\bibitem[{{Ebisawa} {et~al.}(2003){Ebisawa}, {{\.Z}ycki}, {Kubota}, {Mizuno},
  \& {Watarai}}]{Ebisawa2003_kerrd}
{Ebisawa}, K., {{\.Z}ycki}, P., {Kubota}, A., {Mizuno}, T., \& {Watarai}, K.-y.
  2003, \apj, 597, 780, \dodoi{10.1086/378586}

\bibitem[{{Fabian} {et~al.}(2014){Fabian}, {Parker}, {Wilkins}, {Miller},
  {Kara}, {Reynolds}, \& {Dauser}}]{determ_spin_truncation}
{Fabian}, A.~C., {Parker}, M.~L., {Wilkins}, D.~R., {et~al.} 2014, \mnras, 439,
  2307, \dodoi{10.1093/mnras/stu045}

\bibitem[{{Fan} {et~al.}(2024){Fan}, {Li}, {Zhan}, {Liu}, {Zhang}, {Bambi},
  {Ji}, {Ma}, {Steiner}, {Zhang}, \& {Zhou}}]{Fan_maxij1820}
{Fan}, N., {Li}, S., {Zhan}, R., {et~al.} 2024, \apj, 969, 61,
  \dodoi{10.3847/1538-4357/ad49a1}

\bibitem[{{Garc{\'\i}a} {et~al.}(2014){Garc{\'\i}a}, {Dauser}, {Lohfink},
  {Kallman}, {Steiner}, {McClintock}, {Brenneman}, {Wilms}, {Eikmann},
  {Reynolds}, \& {Tombesi}}]{Garcia_relxill}
{Garc{\'\i}a}, J., {Dauser}, T., {Lohfink}, A., {et~al.} 2014, \apj, 782, 76,
  \dodoi{10.1088/0004-637X/782/2/76}

\bibitem[{{Gatuzz} {et~al.}(2019){Gatuzz}, {D{\'\i}az Trigo}, {Miller-Jones},
  \& {Migliari}}]{Gatuzz2019_chandrahighsolution}
{Gatuzz}, E., {D{\'\i}az Trigo}, M., {Miller-Jones}, J.~C.~A., \& {Migliari},
  S. 2019, \mnras, 482, 2597, \dodoi{10.1093/mnras/sty2850}

\bibitem[{{Heil} {et~al.}(2015){Heil}, {Uttley}, \&
  {Klein-Wolt}}]{Heil2015_powercolor}
{Heil}, L.~M., {Uttley}, P., \& {Klein-Wolt}, M. 2015, \mnras, 448, 3339,
  \dodoi{10.1093/mnras/stv191}

\bibitem[{{Homan} \& {Belloni}(2005)}]{Homan_Belloni_2005_statetransitions}
{Homan}, J., \& {Belloni}, T. 2005, \apss, 300, 107,
  \dodoi{10.1007/s10509-005-1197-4}

\bibitem[{Hori {et~al.}(2018)Hori, Ueda, Done, Shidatsu, \&
  Kubota}]{Hori_2018_wind_evolution}
Hori, T., Ueda, Y., Done, C., Shidatsu, M., \& Kubota, A. 2018, The
  Astrophysical Journal, 869, 183, \dodoi{10.3847/1538-4357/aaea5e}

\bibitem[{{Ingram} \& {Motta}(2019)}]{Ingram2019_QPOreview}
{Ingram}, A.~R., \& {Motta}, S.~E. 2019, \nar, 85, 101524,
  \dodoi{10.1016/j.newar.2020.101524}

\bibitem[{{Jones} {et~al.}(1976){Jones}, {Forman}, {Tananbaum}, \&
  {Turner}}]{Jones1976}
{Jones}, C., {Forman}, W., {Tananbaum}, H., \& {Turner}, M.~J.~L. 1976, \apjl,
  210, L9, \dodoi{10.1086/182291}

\bibitem[{{Kallman} \& {Bautista}(2001)}]{Kallman2001_xstar}
{Kallman}, T., \& {Bautista}, M. 2001, \apjs, 133, 221, \dodoi{10.1086/319184}

\bibitem[{{Katoch} {et~al.}(2021){Katoch}, {Baby}, {Nandi}, {Agrawal}, {Antia},
  \& {Mukerjee}}]{Katoch2021_IGRJ17091_GRS1915}
{Katoch}, T., {Baby}, B.~E., {Nandi}, A., {et~al.} 2021, \mnras, 501, 6123,
  \dodoi{10.1093/mnras/staa3756}

\bibitem[{{King} {et~al.}(2014){King}, {Walton}, {Miller}, {Barret}, {Boggs},
  {Christensen}, {Craig}, {Fabian}, {F{\"u}rst}, {Hailey}, {Harrison},
  {Krivonos}, {Mori}, {Natalucci}, {Stern}, {Tomsick}, \&
  {Zhang}}]{kING2014_incl}
{King}, A.~L., {Walton}, D.~J., {Miller}, J.~M., {et~al.} 2014, \apjl, 784, L2,
  \dodoi{10.1088/2041-8205/784/1/L2}

\bibitem[{{Kubota} \& {Makishima}(2004)}]{Kubota_Makishima_color_2004}
{Kubota}, A., \& {Makishima}, K. 2004, \apj, 601, 428, \dodoi{10.1086/380433}

\bibitem[{{Kubota} {et~al.}(1998){Kubota}, {Tanaka}, {Makishima}, {Ueda},
  {Dotani}, {Inoue}, \& {Yamaoka}}]{Kubota1998_colorcorrection}
{Kubota}, A., {Tanaka}, Y., {Makishima}, K., {et~al.} 1998, \pasj, 50, 667,
  \dodoi{10.1093/pasj/50.6.667}

\bibitem[{{Kubota} {et~al.}(2007){Kubota}, {Dotani}, {Cottam}, {Kotani},
  {Done}, {Ueda}, {Fabian}, {Yasuda}, {Takahashi}, {Fukazawa}, {Yamaoka},
  {Makishima}, {Yamada}, {Kohmura}, {Angelini}, \& {Suzaku Team}}]{Kubota2007}
{Kubota}, A., {Dotani}, T., {Cottam}, J., {et~al.} 2007, in IAU Symposium, Vol.
  238, Black Holes from Stars to Galaxies -- Across the Range of Masses, ed.
  V.~{Karas} \& G.~{Matt}, 23--28, \dodoi{10.1017/S1743921307004620}

\bibitem[{{Kushwaha} {et~al.}(2023){Kushwaha}, {Jayasurya}, {Agrawal}, \&
  {Nandi}}]{Kushwaha2023_HSS}
{Kushwaha}, A., {Jayasurya}, K.~M., {Agrawal}, V.~K., \& {Nandi}, A. 2023,
  \mnras, 524, L15, \dodoi{10.1093/mnrasl/slad070}

\bibitem[{{Li} {et~al.}(2005){Li}, {Zimmerman}, {Narayan}, \&
  {McClintock}}]{Li2005_kerrbb}
{Li}, L.-X., {Zimmerman}, E.~R., {Narayan}, R., \& {McClintock}, J.~E. 2005,
  \apjs, 157, 335, \dodoi{10.1086/428089}

\bibitem[{{Lightman} \& {Eardley}(1974)}]{Lightman1974_instability}
{Lightman}, A.~P., \& {Eardley}, D.~M. 1974, \apjl, 187, L1,
  \dodoi{10.1086/181377}

\bibitem[{Liu {et~al.}(2022)Liu, Fu, Bambi, Jiang, Parker, Ji, Kong, Zhang,
  Zhang, \& Zhang}]{Liu2022_1915wind}
Liu, H., Fu, Y., Bambi, C., {et~al.} 2022, Astrophys. J., 933, 122,
  \dodoi{10.3847/1538-4357/ac74b1}

\bibitem[{{Liu} {et~al.}(2023){Liu}, {Bambi}, {Jiang}, {Garc{\'\i}a}, {Ji},
  {Kong}, {Ren}, {Zhang}, \& {Zhang}}]{Liu2023_gx339}
{Liu}, H., {Bambi}, C., {Jiang}, J., {et~al.} 2023, \apj, 950, 5,
  \dodoi{10.3847/1538-4357/acca17}

\bibitem[{{Liu} {et~al.}(2022){Liu}, {Liu}, \& {Bambi}}]{Qichun2022}
{Liu}, Q., {Liu}, H., \& {Bambi}, C. 2022, \mnras, 512, 2082,
  \dodoi{10.1093/mnras/stac616}

\bibitem[{{Lyubarskii}(1997)}]{Lyubarskii1997_accretion_noise}
{Lyubarskii}, Y.~E. 1997, \mnras, 292, 679, \dodoi{10.1093/mnras/292.3.679}

\bibitem[{{Ma} {et~al.}(2021){Ma}, {Tao}, {Zhang}, {Zhang}, {Bu}, {Ge}, {Chen},
  {Qu}, {Zhang}, {Lu}, {Song}, {Yang}, {Yuan}, {Cai}, {Cao}, {Chang}, {Chen},
  {Chen}, {Chen}, {Chen}, {Chen}, {Cui}, {Cui}, {Deng}, {Dong}, {Du}, {Fu},
  {Gao}, {Gao}, {Gao}, {Gu}, {Guan}, {Guo}, {Han}, {Huang}, {Huo}, {Ji}, {Jia},
  {Jiang}, {Jiang}, {Jin}, {Jin}, {Kong}, {Li}, {Li}, {Li}, {Li}, {Li}, {Li},
  {Li}, {Li}, {Li}, {Li}, {Li}, {Liang}, {Liao}, {Liu}, {Liu}, {Liu}, {Liu},
  {Liu}, {Liu}, {Lu}, {Lu}, {Luo}, {Luo}, {Meng}, {Nang}, {Nie}, {Ou}, {Sai},
  {Shang}, {Song}, {Sun}, {Tan}, {Tuo}, {Wang}, {Wang}, {Wang}, {Wang}, {Wang},
  {Wang}, {Wen}, {Wu}, {Wu}, {Wu}, {Xiao}, {Xiao}, {Xie}, {Xiong}, {Xu}, {Xu},
  {Yang}, {Yang}, {Yang}, {Yi}, {Yin}, {You}, {Zhang}, {Zhang}, {Zhang},
  {Zhang}, {Zhang}, {Zhang}, {Zhang}, {Zhang}, {Zhang}, {Zhang}, {Zhang},
  {Zhang}, {Zhang}, {Zhang}, {Zhang}, {Zhang}, {Zhao}, {Zhao}, {Zheng}, {Zhou},
  {Zhou}, {Zhu}, {Zhu}, \& {Zhuang}}]{Ma2021_highenergyQPO}
{Ma}, X., {Tao}, L., {Zhang}, S.-N., {et~al.} 2021, Nature Astronomy, 5, 94,
  \dodoi{10.1038/s41550-020-1192-2}

\bibitem[{{Miller} {et~al.}(2015){Miller}, {Fabian}, {Kaastra}, {Kallman},
  {King}, {Proga}, {Raymond}, \& {Reynolds}}]{Miller_2015}
{Miller}, J.~M., {Fabian}, A.~C., {Kaastra}, J., {et~al.} 2015, \apj, 814, 87,
  \dodoi{10.1088/0004-637X/814/2/87}

\bibitem[{{Miller} {et~al.}(2016){Miller}, {Raymond}, {Fabian}, {Gallo},
  {Kaastra}, {Kallman}, {King}, {Proga}, {Reynolds}, \&
  {Zoghbi}}]{Miller2016_accretionwind1915}
{Miller}, J.~M., {Raymond}, J., {Fabian}, A.~C., {et~al.} 2016, \apjl, 821, L9,
  \dodoi{10.3847/2041-8205/821/1/L9}

\bibitem[{{Mitsuda} {et~al.}(1984){Mitsuda}, {Inoue}, {Koyama}, {Makishima},
  {Matsuoka}, {Ogawara}, {Shibazaki}, {Suzuki}, {Tanaka}, \&
  {Hirano}}]{Mitsuda1984_diskbb}
{Mitsuda}, K., {Inoue}, H., {Koyama}, K., {et~al.} 1984, \pasj, 36, 741

\bibitem[{{Neilsen} {et~al.}(2011){Neilsen}, {Remillard}, \&
  {Lee}}]{Neilsen2011_physicsofheartbeat}
{Neilsen}, J., {Remillard}, R.~A., \& {Lee}, J.~C. 2011, \apj, 737, 69,
  \dodoi{10.1088/0004-637X/737/2/69}

\bibitem[{{Neilsen} {et~al.}(2012){Neilsen}, {Remillard}, \&
  {Lee}}]{Neilson2012_radiationpressure}
---. 2012, \apj, 750, 71, \dodoi{10.1088/0004-637X/750/1/71}

\bibitem[{{Pahari} {et~al.}(2018){Pahari}, {Bhattacharyya}, {Rao},
  {Bhattacharya}, {Vadawale}, {Dewangan}, {McHardy}, {Gandhi}, {Corbel},
  {Schulz}, \& {Altamirano}}]{Pahari2018_spin_wind}
{Pahari}, M., {Bhattacharyya}, S., {Rao}, A.~R., {et~al.} 2018, \apj, 867, 86,
  \dodoi{10.3847/1538-4357/aae53b}

\bibitem[{{Plant} {et~al.}(2014){Plant}, {Fender}, {Ponti}, {Mu{\~n}oz-Darias},
  \& {Coriat}}]{Plant2014_rms}
{Plant}, D.~S., {Fender}, R.~P., {Ponti}, G., {Mu{\~n}oz-Darias}, T., \&
  {Coriat}, M. 2014, \mnras, 442, 1767, \dodoi{10.1093/mnras/stu867}

\bibitem[{{Raj} \& {Nixon}(2021)}]{Raj_Nixon_2021_disktearing}
{Raj}, A., \& {Nixon}, C.~J. 2021, \apj, 909, 82,
  \dodoi{10.3847/1538-4357/abdc25}

\bibitem[{{Ratheesh} {et~al.}(2024){Ratheesh}, {Dov{\v{c}}iak}, {Krawczynski},
  {Podgorn{\'y}}, {Marra}, {Veledina}, {Suleimanov}, {Rodriguez Cavero},
  {Steiner}, {Svoboda}, {Marinucci}, {Bianchi}, {Negro}, {Matt}, {Tombesi},
  {Poutanen}, {Ingram}, {Taverna}, {West}, {Karas}, {Ursini}, {Soffitta},
  {Capitanio}, {Viscolo}, {Manfreda}, {Muleri}, {Parra}, {Beheshtipour},
  {Chun}, {Cibrario}, {Di Lalla}, {Fabiani}, {Hu}, {Kaaret}, {Loktev},
  {Miku{\v{s}}incov{\'a}}, {Mizuno}, {Omodei}, {Petrucci}, {Puccetti},
  {Rankin}, {Zane}, {Zhang}, {Agudo}, {Antonelli}, {Bachetti}, {Baldini},
  {Baumgartner}, {Bellazzini}, {Bongiorno}, {Bonino}, {Brez}, {Bucciantini},
  {Castellano}, {Cavazzuti}, {Chen}, {Ciprini}, {Costa}, {De Rosa}, {Del
  Monte}, {Di Gesu}, {Di Marco}, {Donnarumma}, {Doroshenko}, {Ehlert}, {Enoto},
  {Evangelista}, {Ferrazzoli}, {Garcia}, {Gunji}, {Hayashida}, {Heyl},
  {Iwakiri}, {Jorstad}, {Kislat}, {Kitaguchi}, {Kolodziejczak}, {La Monaca},
  {Latronico}, {Liodakis}, {Maldera}, {Marin}, {Marscher}, {Marshall},
  {Massaro}, {Mitsuishi}, {Ng}, {O'Dell}, {Oppedisano}, {Papitto}, {Pavlov},
  {Peirson}, {Perri}, {Pesce-Rollins}, {Pilia}, {Possenti}, {Ramsey},
  {Roberts}, {Romani}, {Sgr{\`o}}, {Slane}, {Spandre}, {Swartz}, {Tamagawa},
  {Tavecchio}, {Tawara}, {Tennant}, {Thomas}, {Trois}, {Tsygankov}, {Turolla},
  {Vink}, {Weisskopf}, {Wu}, \& {Xie}}]{Ratheesh2024_HSS}
{Ratheesh}, A., {Dov{\v{c}}iak}, M., {Krawczynski}, H., {et~al.} 2024, \apj,
  964, 77, \dodoi{10.3847/1538-4357/ad226e}

\bibitem[{{Rawat} {et~al.}(2022){Rawat}, {Misra}, {Jain}, \&
  {Yadav}}]{Rawat2022_GRS1915_timeresolved}
{Rawat}, D., {Misra}, R., {Jain}, P., \& {Yadav}, J.~S. 2022, \mnras, 511,
  1841, \dodoi{10.1093/mnras/stac154}

\bibitem[{{Remillard} {et~al.}(1999){Remillard}, {Morgan}, {McClintock},
  {Bailyn}, \& {Orosz}}]{Remillard1999_GROJ1655}
{Remillard}, R.~A., {Morgan}, E.~H., {McClintock}, J.~E., {Bailyn}, C.~D., \&
  {Orosz}, J.~A. 1999, \apj, 522, 397, \dodoi{10.1086/307606}

\bibitem[{{Remillard} {et~al.}(2022){Remillard}, {Loewenstein}, {Steiner},
  {Prigozhin}, {LaMarr}, {Enoto}, {Gendreau}, {Arzoumanian}, {Markwardt},
  {Basak}, {Stevens}, {Ray}, {Altamirano}, \& {Buisson}}]{Remillard2022_3C50}
{Remillard}, R.~A., {Loewenstein}, M., {Steiner}, J.~F., {et~al.} 2022, \aj,
  163, 130, \dodoi{10.3847/1538-3881/ac4ae6}

\bibitem[{{Seifina} {et~al.}(2014){Seifina}, {Titarchuk}, \&
  {Shaposhnikov}}]{Seifina2014_mass}
{Seifina}, E., {Titarchuk}, L., \& {Shaposhnikov}, N. 2014, \apj, 789, 57,
  \dodoi{10.1088/0004-637X/789/1/57}

\bibitem[{{Shimura} \& {Takahara}(1995)}]{color_correction}
{Shimura}, T., \& {Takahara}, F. 1995, \apj, 445, 780, \dodoi{10.1086/175740}

\bibitem[{{Sriram} {et~al.}(2016){Sriram}, {Rao}, \& {Choi}}]{Sriram}
{Sriram}, K., {Rao}, A.~R., \& {Choi}, C.~S. 2016, \apj, 823, 67,
  \dodoi{10.3847/0004-637X/823/1/67}

\bibitem[{{Steiner} {et~al.}(2010){Steiner}, {McClintock}, {Remillard}, {Gou},
  {Yamada}, \& {Narayan}}]{Jack2010_LMCX3}
{Steiner}, J.~F., {McClintock}, J.~E., {Remillard}, R.~A., {et~al.} 2010,
  \apjl, 718, L117, \dodoi{10.1088/2041-8205/718/2/L117}

\bibitem[{{Stiele} \& {Kong}(2020)}]{Stiele2020_1820HRD}
{Stiele}, H., \& {Kong}, A.~K.~H. 2020, \apj, 889, 142,
  \dodoi{10.3847/1538-4357/ab64ef}

\bibitem[{{Swank}(2001)}]{Swank2001}
{Swank}, J. 2001, Astrophysics and Space Science Supplement, 276, 201,
  \dodoi{10.1023/A:1011659800688}

\bibitem[{{Trudolyubov} {et~al.}(2001){Trudolyubov}, {Borozdin}, \&
  {Priedhorsky}}]{Trudolyubov2001_1998qrm}
{Trudolyubov}, S.~P., {Borozdin}, K.~N., \& {Priedhorsky}, W.~C. 2001, \mnras,
  322, 309, \dodoi{10.1046/j.1365-8711.2001.04073.x}

\bibitem[{{Trueba} {et~al.}(2019){Trueba}, {Miller}, {Kaastra}, {Zoghbi},
  {Fabian}, {Kallman}, {Proga}, \& {Raymond}}]{Trueba2019_4U1630wind}
{Trueba}, N., {Miller}, J.~M., {Kaastra}, J., {et~al.} 2019, \apj, 886, 104,
  \dodoi{10.3847/1538-4357/ab4f70}

\bibitem[{{Uttley} \& {Malzac}(2024)}]{2024Uttley_Malzac}
{Uttley}, P., \& {Malzac}, J. 2024, \mnras, \dodoi{10.1093/mnras/stae2514}

\bibitem[{{Uttley} {et~al.}(2011){Uttley}, {Wilkinson}, {Cassatella}, {Wilms},
  {Pottschmidt}, {Hanke}, \& {B{\"o}ck}}]{Uttley2011_hardlag_GX339}
{Uttley}, P., {Wilkinson}, T., {Cassatella}, P., {et~al.} 2011, \mnras, 414,
  L60, \dodoi{10.1111/j.1745-3933.2011.01056.x}

\bibitem[{{van der Klis}(1995)}]{vanderklis1995}
{van der Klis}, M. 1995, in X-ray Binaries, ed. W.~H.~G. {Lewin}, J.~{van
  Paradijs}, \& E.~P.~J. {van den Heuvel}, 252--307

\bibitem[{{Verner} {et~al.}(1996){Verner}, {Ferland}, {Korista}, \&
  {Yakovlev}}]{Verner_crosssection}
{Verner}, D.~A., {Ferland}, G.~J., {Korista}, K.~T., \& {Yakovlev}, D.~G. 1996,
  \apj, 465, 487, \dodoi{10.1086/177435}

\bibitem[{{Wang} {et~al.}(2021){Wang}, {Mastroserio}, {Kara}, {Garc{\'\i}a},
  {Ingram}, {Connors}, {van der Klis}, {Dauser}, {Steiner}, {Buisson}, {Homan},
  {Lucchini}, {Fabian}, {Bright}, {Fender}, {Cackett}, \&
  {Remillard}}]{Jingyi_2021_diskcoronajetconnection}
{Wang}, J., {Mastroserio}, G., {Kara}, E., {et~al.} 2021, \apjl, 910, L3,
  \dodoi{10.3847/2041-8213/abec79}

\bibitem[{{Wang} {et~al.}(2024){Wang}, {Kara}, {Garc{\'\i}a}, {Altamirano},
  {Belloni}, {Steiner}, {van der Klis}, {Ingram}, {Mastroserio}, {Connors},
  {Lucchini}, {Dauser}, {Neilsen}, {Lewin}, {Remillard}, \&
  {Homan}}]{Wang2024_IGRJ17091}
{Wang}, J., {Kara}, E., {Garc{\'\i}a}, J.~A., {et~al.} 2024, \apj, 963, 14,
  \dodoi{10.3847/1538-4357/ad1595}

\bibitem[{{Wang-Ji} {et~al.}(2018){Wang-Ji}, {Garc{\'\i}a}, {Steiner},
  {Tomsick}, {Harrison}, {Bambi}, {Petrucci}, {Ferreira}, {Chakravorty}, \&
  {Clavel}}]{Wang2018_gx339}
{Wang-Ji}, J., {Garc{\'\i}a}, J.~A., {Steiner}, J.~F., {et~al.} 2018, \apj,
  855, 61, \dodoi{10.3847/1538-4357/aaa974}

\bibitem[{Wilms {et~al.}(2000)Wilms, Allen, \& McCray}]{Wilms_tbabs}
Wilms, J., Allen, A., \& McCray, R. 2000, The Astrophysical Journal, 542, 914,
  \dodoi{10.1086/317016}

\bibitem[{Yan {et~al.}(2018)Yan, Ji, Liu, M{\'e}ndez, Wang, Li, Qu, Sun, Ge,
  Liao, {et~al.}}]{yan2018corona_disc_1915}
Yan, S.-P., Ji, L., Liu, S.-M., {et~al.} 2018, Monthly Notices of the Royal
  Astronomical Society, 474, 1214

\bibitem[{{Yang} {et~al.}(2022){Yang}, {Zhang}, {Huang}, {Bu}, {Zhang}, {Liu},
  {Yu}, {Wang}, {Zhao}, {Tao}, {Qu}, {Zhang}, {Zhang}, {Song}, {Lu}, {Cao},
  {Chen}, {Cai}, {Chang}, {Chen}, {Chen}, {Chen}, {Chen}, {Cui}, {Ding}, {Du},
  {Gao}, {Gao}, {Ge}, {Gu}, {Guan}, {Guo}, {Han}, {Huo}, {Jia}, {Jiang}, {Jin},
  {Kong}, {Li}, {Li}, {Li}, {Li}, {Li}, {Li}, {Li}, {Lin}, {Liu}, {Li}, {Li},
  {Liang}, {Liao}, {Liu}, {Liu}, {Lu}, {Luo}, {Luo}, {Ma}, {Ma}, {Ma}, {Meng},
  {Nang}, {Nie}, {Ou}, {Ren}, {Sai}, {Song}, {Sun}, {Tan}, {Tuo}, {Wang},
  {Wang}, {Wang}, {Wang}, {Wang}, {Wen}, {Wu}, {Wu}, {Wu}, {Xiao}, {Xu},
  {Xiong}, {Yang}, {Yang}, {Yi}, {Yin}, {You}, {Zhang}, {Zhang}, {Zhang},
  {Zhang}, {Zhang}, {Zhang}, {Zhang}, {Zhang}, {Zhang}, {Zhang}, {Zhao},
  {Zhao}, {Zheng}, \& {Zhou}}]{Yang2022_qrm2021}
{Yang}, Z.-x., {Zhang}, L., {Huang}, Y., {et~al.} 2022, \apj, 937, 33,
  \dodoi{10.3847/1538-4357/ac84d6}

\bibitem[{{Zdziarski} {et~al.}(2020){Zdziarski}, {Szanecki}, {Poutanen},
  {Gierli{\'n}ski}, \& {Biernacki}}]{Zdziarski2020_thcomp}
{Zdziarski}, A.~A., {Szanecki}, M., {Poutanen}, J., {Gierli{\'n}ski}, M., \&
  {Biernacki}, P. 2020, \mnras, 492, 5234, \dodoi{10.1093/mnras/staa159}

\bibitem[{{Zhang} {et~al.}(2024){Zhang}, {Bambi}, {Liu}, {Jiang}, {Shi},
  {Zhang}, {Young}, {Tomsick}, {Coughenour}, \& {Zhou}}]{Zhang2024_1803wind}
{Zhang}, Z., {Bambi}, C., {Liu}, H., {et~al.} 2024, \apj, 975, 22,
  \dodoi{10.3847/1538-4357/ad7b29}

\bibitem[{{Zoghbi} {et~al.}(2016){Zoghbi}, {Miller}, {King}, {Miller}, {Proga},
  {Kallman}, {Fabian}, {Harrison}, {Kaastra}, {Raymond}, {Reynolds}, {Boggs},
  {Christensen}, {Craig}, {Hailey}, {Stern}, \&
  {Zhang}}]{Zoghbi2016_diskwindconnection1915}
{Zoghbi}, A., {Miller}, J.~M., {King}, A.~L., {et~al.} 2016, \apj, 833, 165,
  \dodoi{10.3847/1538-4357/833/2/165}

\bibitem[{{{\.Z}ycki} {et~al.}(1999){{\.Z}ycki}, {Done}, \&
  {Smith}}]{Zycki1999}
{{\.Z}ycki}, P.~T., {Done}, C., \& {Smith}, D.~A. 1999, \mnras, 309, 561,
  \dodoi{10.1046/j.1365-8711.1999.02885.x}

\end{thebibliography}

\end{document}